\documentclass[reprint,twocolumn,pre,showpacs,amsmath,amssymb,aps]{revtex4-2}

\usepackage{natbib}
\usepackage{graphicx,color,rotating}
\usepackage[latin1]{inputenc}
\usepackage{textcomp}
\usepackage{dcolumn}
\usepackage{bm}     
\usepackage{upgreek}
\usepackage{subdepth}  			
\usepackage{epstopdf}   		
\usepackage[latin1]{inputenc}
\usepackage{amsmath}
\usepackage{amssymb}
\usepackage{amsfonts}
\usepackage{array}
\newcolumntype{L}[1]{>{\raggedright\let\newline\\\arraybackslash\hspace{0pt}}m{#1}}
\newcolumntype{C}[1]{>{\centering\let\newline\\\arraybackslash\hspace{0pt}}m{#1}}
\newcolumntype{R}[1]{>{\raggedleft\let\newline\\\arraybackslash\hspace{0pt}}m{#1}}

\usepackage{hyperref}					
\usepackage{xcolor}						
\hypersetup{			     			
    colorlinks,
    allcolors={blue}
}
\newcommand{\qmarks}[1]{{``#1''}}
\newcommand{\qmarkstt}[1]{{``\texttt{#1}''}}

\newcommand{\mr}[1]{\ensuremath{\mathrm{#1}}}
\newcommand{\myvec}[1]{\bm{#1}}
\newcommand{\ee}{\mathrm{e}}
\newcommand{\ii}{\mathrm{i}}
\newcommand{\dm}{\mathrm{d}}

\newcommand{\avr}[1]{\big\langle #1 \big\rangle}
\newcommand{\abs}[1]{\big|{#1}\big|}
\newcommand{\intI}[2]{I_{{#1}{#2}}^{(1)}}
\newcommand{\intII}[2]{I_{{#1}{#2}}^{(2)}}

\newcommand{\intn}[2]{I_{{#1}{#2}}^{(n)}}

\DeclareMathOperator{\re}{Re}

\newcommand{\iot}{{\ii\omega t}}

\newcommand{\pp}{\partial}
\newcommand{\ppsqr}{\partial^{\,2_{}}}

\newcommand{\nablabf}{\boldsymbol{\nabla}}

\newcommand{\divop}{\nablabf\cdot}

\newcommand{\scap}{\!\cdot\!}

\newcommand{\AAA}{\myvec{A}}

\newcommand{\BBB}{\myvec{B}}

\newcommand{\eee}{\myvec{e}}
\newcommand{\een}{\myvec{e}}

\newcommand{\fff}{\myvec{f}}

\newcommand{\fffac}{\fff_\mathrm{ac}}

\newcommand{\III}{\myvec{I}}

\newcommand{\kc}{k_\mathrm{c}}
\newcommand{\kt}{k_\mathrm{t}}
\newcommand{\ks}{k_\mathrm{s}}
\newcommand{\kcsqr}{k^{2_{}}_\mathrm{c}}

\newcommand{\nnn}{\myvec{n}}

\newcommand{\sss}{\myvec{s}}

\newcommand{\uuu}{\myvec{u}}

\newcommand{\VVV}{\myvec{V}}

\newcommand{\vvv}{\myvec{v}}

\newcommand{\vvvIdO}{\myvec{v}^{d0}_1}

\newcommand{\vvvIdel}{\myvec{v}^\delta_1}
\newcommand{\vvvIdelO}{\myvec{v}^{\delta0}_1}

\newcommand{\zerovec}{\boldsymbol{0}}

\newcommand{\calL}{\mathcal{L}}

\renewcommand{\cp}{c_p}
\newcommand{\cpO}{c_{p0}}

\newcommand{\Dth}{D^\mathrm{th}}
\newcommand{\DthO}{D^\mathrm{th}_0}

\newcommand{\Eac}{E_\mathrm{ac}}
\newcommand{\Lac}{\calL_\mathrm{ac}}

\newcommand{\Pac}{P_\mathrm{ac}}

\newcommand{\kth}{k^\mathrm{th}}

\newcommand{\kthO}{k^\mathrm{th}_0}
\newcommand{\kthOsl}{k^\mathrm{th,sl}_0}
\newcommand{\kthOfl}{k^\mathrm{th,fl}_0}
\newcommand{\kthsl}{k^\mathrm{th,sl}}
\newcommand{\kthfl}{k^\mathrm{th,fl}}
\newcommand{\kthI}{k^\mathrm{th}_1}
\newcommand{\kthIdelta}{k^{\mathrm{th},\delta}_1}
\newcommand{\kthIdeltaO}{k^{\mathrm{th},\delta0}_1}
\newcommand{\kthId}{k^{\mathrm{th},d}_1}
\newcommand{\kthIdO}{k^{\mathrm{th},d0}_1}

\newcommand{\kthIsl}{k^\mathrm{th,sl}_1}
\newcommand{\kthIfl}{k^\mathrm{th,fl}_1}

\newcommand{\kapT}{\kappa_T}

\newcommand{\kapS}{\kappa_s}

\newcommand{\kaps}{\kappa_s}
\newcommand{\kapsO}{\kappa_{s0}}

\newcommand{\vbc}{v_\mathrm{bc}}

\newcommand{\alphapO}{{\alpha_{p0}}}
\newcommand{\alfP}{{\alpha_p}}

\newcommand{\delt}{\delta_t}
\newcommand{\deltInv}{\delta_t^{-1}}

\newcommand{\deltsqrInv}{\delta_t^{-2}}
\newcommand{\dels}{\delta_s}
\newcommand{\delsInv}{\delta_s^{-1}}

\newcommand{\delssqrInv}{\delta_s^{-2}}

\newcommand{\etaB}{\eta^\mathrm{b}}

\newcommand{\etaO}{\eta_0}

\newcommand{\etaI}{\eta_1}
\newcommand{\etaId}{\eta_1^d}
\newcommand{\etaIdelta}{\eta_1^\delta}

\newcommand{\Gams}{\Gamma_\mathrm{s}}
\newcommand{\GamOcfl}{\Gamma^\mr{fl}_\mathrm{0c}}

\newcommand{\cs}{c_s}

\newcommand{\kOsqr}{k^{2_{}}_0}

\newcommand{\pI}{p_1}

\newcommand{\taubfI}{\bm{\tau}_1}

\newcommand{\TOd}{T^d_0}
\newcommand{\TOdel}{T_0^\delta}
\newcommand{\TOdelO}{T_0^{\delta0}}
\newcommand{\TOfl}{T_0^{\mr{fl}}}
\newcommand{\TOfld}{T_0^{\mr{fl},d}}
\newcommand{\TOfldO}{T_0^{\mr{fl},d0}}
\newcommand{\TOfldel}{T_0^{\mr{fl},\delta}}
\newcommand{\TOfldelO}{T_0^{\mr{fl},\delta0}}
\newcommand{\TOsl}{T_0^{\mr{sl}}}
\newcommand{\TOsld}{T_0^{\mr{sl},d}}
\newcommand{\TOsldO}{T_0^{\mr{sl},d0}}
\newcommand{\TOsldel}{T_0^{\mr{sl},\delta}}
\newcommand{\TOsldelO}{T_0^{\mr{sl},\delta0}}
\newcommand{\TId}{T_1^d}
\newcommand{\TIdel}{T_1^\delta}
\newcommand{\TIdelS}{T_1^{\delta*}}
\newcommand{\TIdelO}{T_1^{\delta0}}
\newcommand{\TIdelOS}{T_1^{\delta0*}}
\newcommand{\TIfl}{T_1^{\mr{fl}}}
\newcommand{\TIfld}{T_1^{\mr{fl},d}}
\newcommand{\TIfldS}{T_1^{\mr{fl},d*}}
\newcommand{\TIfldO}{T_1^{\mr{fl},d0}}
\newcommand{\TIfldel}{T_1^{\delta,\mr{fl}}}
\newcommand{\TIfldelS}{T_1^{\delta*,\mr{fl}}}
\newcommand{\TIfldelO}{T_1^{\delta0,\mr{fl}}}
\newcommand{\TIfldelOS}{T_1^{\delta0*,\mr{fl}}}
\newcommand{\TIsl}{T_1^{\mr{sl}}}
\newcommand{\TIsld}{T_1^{\mr{sl},d}}
\newcommand{\TIsldO}{T_1^{\mr{sl},d0}}
\newcommand{\TIsldel}{T_1^{\delta,\mr{sl}}}
\newcommand{\TIsldelO}{T_1^{\delta0,\mr{sl}}}
\newcommand{\TO}{T_0}
\newcommand{\TI}{T_1}

\newcommand{\vvvO}{\vvv_0}

\newcommand{\vvvI}{\vvv_1}

\newcommand{\rhoOfl}{\rho^\fl_0}
\newcommand{\rhoOsl}{\rho^\sl_0}
\newcommand{\rhoO}{\rho_0}

\newcommand{\rhoI}{\rho_1}

\newcommand{\SIC}{\textrm{C}}
\newcommand{\SICel}{^\circ\!\textrm{C}}

\newcommand{\SIum}{\upmu\textrm{m}}

\newcommand{\SIMHz}{\textrm{MHz}}

\newcommand{\SIJ}{\textrm{J}}

\newcommand{\SIK}{\textrm{K}}
\newcommand{\SImK}{\textrm{mK}}

\newcommand{\SIm}{\textrm{m}}

\newcommand{\SImm}{\textrm{mm}}
\newcommand{\SImum}{\textrm{\textmu{}m}}

\newcommand{\SInm}{\textrm{nm}}

\newcommand{\SIMPa}{\textrm{MPa}}

\newcommand{\SIs}{\textrm{s}}
\newcommand{\SIms}{\textrm{ms}}

\newcommand{\SIns}{\textrm{ns}}

\newcommand{\SIV}{\textrm{V}}

\newcommand{\SImW}{\textrm{mW}}
\newcommand{\nn}{\nonumber}
\newcommand{\beq}[1]{\begin{equation} \eqlab{#1}}
\newcommand{\eeq}{\end{equation}}
\newcommand{\bsub}{\begin{subequations}}
\newcommand{\esub}{\end{subequations}}
\def\bal#1\eal{\begin{align}#1\end{align}}
\def\balat#1#2\ealat{\begin{alignat}{#1} #2 \end{alignat}}
\def\bsubal#1 #2\esubal{\bsuba{#1}\begin{align}#2\end{align} \esuba}     
\def\bsubalat#1#2#3\esubalat{\bsuba{#1} \begin{alignat}{#2} #3 \end{alignat} \esuba}
\newcommand{\bsuba}[1]{\bsub \eqlab{#1}}
\newcommand{\esuba}{\esub}

\newcommand{\eqlab}[1]{\label{eq:#1}}
\renewcommand{\eqref}[1]{Eq.~(\ref{eq:#1})}
\newcommand{\eqnoref}[1]{(\ref{eq:#1})}

\newcommand{\eqsref}[2]{Eqs.~(\ref{eq:#1}) and~(\ref{eq:#2})}
\newcommand{\eqsnoref}[2]{(\ref{eq:#1}) and~(\ref{eq:#2})}

\newcommand{\figref}[1]{Fig.~\ref{fig:#1}}

\newcommand{\figlab}[1]{\label{fig:#1}}
\newcommand{\appref}[1]{Appendix~\ref{sec:#1}}

\newcommand{\secref}[1]{Section~\ref{sec:#1}}

\newcommand{\seclab}[1]{\label{sec:#1}}

\newcommand{\sigmabf}{\bm{\sigma}}

\newcommand{\taubf}{\bm{\tau}}

\newcommand{\cL}{c_\mathrm{lo}}
\newcommand{\cT}{c_\mathrm{tr}}

\newcommand{\cLsqr}{c^2_\mathrm{lo}}
\newcommand{\cTsqr}{c^2_\mathrm{tr}}

\newcommand{\fl}{\mathrm{fl}}
\newcommand{\xl}{\mathrm{xl}}
\renewcommand{\sl}{\mathrm{sl}}

\definecolor{darkgreen}{rgb}{0.00, 0.50, 0.00}
\definecolor{DARKGREEN}{rgb}{0.00, 0.50, 0.00}
\definecolor{RED}{rgb}{1.00, 0.00, 0.00}
\definecolor{GREEN}{rgb}{0.00, 1.00, 0.00}
\definecolor{BLUE}{rgb}{0.00, 0.00, 1.00}
\definecolor{MAGENTA}{rgb}{1.00, 0.00, 1.00}

\begin{document}


\title{Theory and modeling of nonperturbative effects at high acoustic energy densities in thermoviscous acoustofluidics}

\author{Jonas Helboe Joergensen}
\email{jonashj@fysik.dtu.dk}
\affiliation{Department of Physics, Technical University of Denmark,\\ DTU Physics Building 309, DK-2800 Kongens Lyngby, Denmark}

\author{Henrik Bruus}
\email{bruus@fysik.dtu.dk}
\affiliation{Department of Physics, Technical University of Denmark,\\
DTU Physics Building 309, DK-2800 Kongens Lyngby, Denmark}

\date{20 December 2021}

\begin{abstract}
A theoretical model of thermal boundary layers and acoustic heating in microscale acoustofluidic devices is presented. It includes effective boundary conditions allowing for simulations in three dimensions. The model is extended by an iterative scheme to incorporate nonlinear thermoviscous effects not captured by standard perturbation theory. The model predicts that the dominant nonperturbative effects in these devices are due to the dependency of thermoacoustic streaming on gradients in the steady temperature induced by a combination of internal frictional heating, external heating, and thermal convection. The model enables simulations in a nonperturbative regime relevant for design and fabrication of high-throughput acoustofluidic devices.
\end{abstract}
\maketitle

\section{Introduction}
\seclab{intro}
Modeling and simulations of acoustofluidic devices is used to optimize and develop designs of microscale acoustofluidic devices. Traditionally, acoustofluidic models have been based on perturbation theory \cite{Muller2012, Muller2014, Hahn2015, Hahn2015a, Bach2018, Skov2019, Joergensen2021}, but in this paper we present an iterative numerical model that enables simulations of nonlinear acoustofluidics beyond the amplitude limitations set by perturbation theory. The motivation for this theoretical development is that combined theoretical and experimental studies of acoustic streaming has shown that the perturbative treatment is pushed to its limit and beyond \cite{Qiu2021}. In the latter work, the perturbation model is challenged by the fast acoustic streaming, which creates a significant convection, and, as discussed below, also heating from friction in the viscous boundary layers can be important. Both of these effects are not described by standard perturbation theory.

The validity of perturbation theory is mainly challenged in systems with a high acoustic energy density $\Eac$, needed in the development of faster acoustofluidic handling of suspended particles and molecular suspensions. In particular, the volumetric throughput is often a limiting factor for clinical use of acoustofluidic devices \cite{Adams2012, Chen2016, Antfolk2017, Wu2019}, so it is of general interest to develop a model that allows for simulation of such high-$\Eac$ devices. Nonlinear effects due to fast acoustic streaming have previously been studied numerically in gases with a model using an ideal analytical pressure field \cite{michel2019}. Those models include the nonlinear effects of changing the temperature field by convection due to the streaming velocity field, but the nonlinear effect of acoustic heating and nonlinearities in the acoustic fields themselves are not included.

Numerical models in acoustofluidics can be categorized as viscous models \cite{Muller2012, Bach2018} and thermoviscous models \cite{Muller2014, Joergensen2021}, as well as \ full models \cite{Muller2012, Muller2014} and effective models \cite{Bach2018, Joergensen2021}. The viscous models include the full viscous fluid description, but assume an adiabatic temperature field governed by the pressure field and typically assume temperature-independent material parameters. The thermoviscous models further include thermal boundary layers, temperature-dependent material parameters, and heating created in the viscous boundary layers. Full models require numerical resolution of the thin boundary layers, and they are therefore numerically expensive. In  contrast, effective models include analytical expressions for the boundary layers, so a fine boundary-layer mesh is avoided, and they therefore enables three-dimensional (3D) simulations. In this work we build on and expand our previous perturbative thermoviscous effective model \cite{Joergensen2021} by including frictional heating due to the acoustic field, and by going beyond perturbation theory with the introduction of an iterative scheme including quasi-steady and acoustic fields to allow for higher acoustic energy densities and larger thermal convection. The developed  nonlinear, effective-boundary-layer, thermoviscous acoustofluidic model is the main result of this work. Experimental validation of the model is presented in a concurrent Letter \cite{Joergensen2022}.

In \secref{Model} the basic assumptions and governing equations for the nonperturbative appraoch are presented. In \secref{AcousticFields}, our perturbative theory \cite{Joergensen2021} of the thermoviscous acoustic fields is briefly summarized. In \secref{SteadyFields}, the known theory  \cite{Joergensen2021} for the steady mechanical fields is presented, and we develop the theory for the steady temperature fields. In \secref{iterative}, we present the nonperturbative iterative procedure for computing nonlinear thermoacoustic effects, and we briefly explain how to implement it in the software Comsol Multiphysics \cite{Comsol56}. In \secref{ModelValidation}, we validate the implementation of our numerical model, and we present two model examples of nonlinear acoustofluidics: a model in two dimensions (2D) involving internal acoustic heating, and a model in three dimensions (3D) involving thermoacoustic streaming induced by absorption of light. Finally, we conclude in \secref{conclusions}.

\section{Theory and model assumptions}
\seclab{Model}
Based on our previous perturbative approach \cite{Joergensen2021}, we consider an acoustofluidic device consisting of an elastic solid containing a microchannel filled with a thermoviscous Newtonian fluid and actuated by a piezoelectric transducer at a single frequency in the $\SIMHz$ range. Due to the internal dissipation and hydrodynamic nonlinearities in the fluid, the resulting time-harmonic acoustic field leads to a time-averaged response in the form of acoustic streaming and steady temperature gradients.  For simplicity, the piezoelectric transducer is left out of the analysis, and is only represented by an oscillating displacement condition on part of the surface of the elastic solid. We have in other studies included a full model of the transducer in the numerical model \cite{Skov2019, Lickert2021, Bode2021}.

\subsection{Governing equations}
The response of the fluid embedded in the elastic solid to the oscillating-displacement boundary condition is controlled by the hydro-, elasto-, and thermodynamic governing equations of the coupled thermoviscous fluid and elastic solid. The linear elastic solid is described in the Lagrangian picture by the fields of the density $\rho$, the displacement $\uuu$, the temperature $T$, and by the following material parameters: the longitudinal and transverse sound speeds $\cL$ and $\cT$, the thermal conductivity $\kth$, the specific heat capacity $\cp$, the ratio of specific heat capacities  $\gamma = c_p/c_v$, the thermal expansion coefficient $\alfP$, and the isentropic and isothermal compressibilities $\kapS$ and $\kapT$. The velocity field is the time derivative of the displacement field, $\vvv^\mr{sl} =\pp_t \uuu$, so no advection occur, and the governing equations are the transport equations of the momentum density $\rho\pp_t\uuu$ and temperature $T$ \cite{Landau1986, Karlsen2015},
 \bsubal{govEQ_sl}
 \eqlab{govU_sl}
 \rho  \ppsqr_t\uuu & = \nablabf \cdot \sigmabf,
 \\
 \eqlab{govT_sl}
 \pp_t T + \frac{(\gamma-1)}{\alfP}\pp_t (\nablabf \cdot \uuu)
 &= \frac{\gamma}{\rho\cp}\nablabf \cdot (\kth\nablabf T) + P,
 \esubal
where $P$ is the external heat power density, and $\sigmabf$ is the stress tensor, which for isotropic solids is,
\bsubal{stress_sl}
 \sigmabf &= -\frac{\alfP}{\kapT} (T-\TO) \III + \taubf,
 \\
 \taubf   &= \rho\cTsqr \Big[\nablabf\uuu + (\nablabf\uuu)^\dagger\Big]
             + \rho\big(\cLsqr-2\cTsqr\big)(\nablabf \cdot \uuu) \III.
\esubal

The fluid is described in the Eulerian picture by the fields of the density $\rho$, the pressure $p$, the velocity $\vvv$, the temperature $T$, and the energy per mass unit $\epsilon$, and by the following material parameters: the dynamic and bulk viscosity $\eta$ and $\etaB$, the thermal conductivity $\kth$, the specific heat $\cp$, the thermal expansion coefficient $\alfP$, the ratio of specific heats  $\gamma = c_p/c_v$, and the isentropic and isothermal compressibilities $\kapS$ and $\kapT=\gamma \kapS$. The governing equations are the transport equations for the density of mass $\rho$, momentum $\rho\vvv$, and internal energy~$\rho\epsilon$, \cite{Landau1993, Muller2014, Karlsen2015}
 \bsubal{govEQ_fl}
 \eqlab{govRho_fl}
 \pp_t \rho &= -\nablabf\scap(\rho \vvv),
 \\
 \eqlab{govV_fl}
 \pp_t(\rho \vvv) &= \nablabf \scap (\sigmabf -\rho \vvv \vvv),
 \\
 \eqlab{govT_fl}
 \pp_t\Big( \rho \epsilon + \rho \frac{v^2}{2} \Big)
 &= \nablabf\scap \Big[\kth \nablabf T +\vvv\scap\sigmabf
 - \rho \vvv \Big(\epsilon + \frac{v^2}{2}  \Big) \Big] + P,
 \esubal
where $P$ is the external heat power density, and $\sigmabf$ is the stress tensor,
 \bsubal{stress_fl}
 \sigmabf &= -p \III + \taubf,
 \\
 \taubf &= \eta \Big[\nablabf \vvv +(\nablabf \vvv)^\dagger\Big]
 +\Big(\etaB - \frac23 \eta\Big)(\nablabf \cdot \vvv) \III.
 \esubal

Pressure $p$ and temperature $T$ are related to the internal energy density $\epsilon$ by the first law of thermodynamics, and to the density $\rho$ by the equation of state \cite{Landau1980, Muller2014, Karlsen2015},
 \bsubal{thermodynamics}
 \eqlab{thermoFirstLaw}
 \rho \dm\epsilon &= (\rho\cp -\alfP p)\: \dm T + (\kapT p-\alfP T)\: \dm p
 \\
 \eqlab{eqofstate}
 \dm \rho &= - \rho\alfP\: dT + \rho\kapT\: \dm p
 \esubal
Like the density, any material parameter $q$ has a temperature and pressure dependency,
\bsub
 \bal\eqlab{water_property}
 \frac{1}{q_0}\:\dm q & = a_q^T\: \alfP\: \dm T + a_q^p\:\kapT\:\dm p,
 \\
 a_q^T & = \frac{1}{\alfP q_0} \Big(\frac{\pp q}{\pp T}\Big)_p, \quad
 a_q^p = \frac{1}{\kapT q_0} \Big(\frac{\pp q}{\pp p}\Big)_T.
 \eal
Note that here the variables are $(T, p)$ and not $(T,\rho)$ as in Refs.~\cite{Muller2014, Joergensen2021}. For a pressure change $\dm p$ accompanied by an adiabatic temperature change $\dm T = (\gamma-1)\frac{\kapT}{\alfP} \dm p$, the adiabatic pressure dependency of a parameter $q$ is,
 \eqlab{water_property_adiabatic}
 \bal
 \frac{1}{q_0}\:\dm q & = a_q^T\: \alfP\: \dm T + a_q^p\:\kapT\:\dm p
 = a_q^{p,\mr{ad}} \kaps p_1,
 \\
 a_q^{p,\mr{ad}} & = \gamma\left(\gamma-1\right)\: a_q^T\:  +\gamma a_q^p.
 \eal
 \esub
For steady temperature gradients and oscillating thermal boundary layers, $a_q^T$ is relevant and for bulk adiabatic pressure wave $a_q^{p,\mr{ad}}$ is the relevant quantity.
For water at $T=25~\SICel$, using \eqsref{eqofstate}{water_property_adiabatic}, we compute  the dimensionless sensitivities $a_q^T$, $a_q^p$ and $a_q^{p,\mr{ad}}$ from the $T$-$\rho$ dependencies of the parameters $q$ listed in Ref.~\cite{Muller2014},
 \beq{a_water}
 \begin{array}{rcrrcr}
 a^T_\rho    &=&   -1, & \qquad
  a^T_{\kapS} &=&  -10,
\\[1mm]
 a^p_\rho    &=&   1, & \qquad
  a^{p,\mr{ad}}_{\rho} &=&  1,
  \\[1mm]
 a^T_\eta    &=&  -88, & \qquad
  a^{p,\mr{ad}}_\eta    &=&  -1.3,
\\[1mm]
 a^T_{\etaB} &=& -100, &\qquad
  a^{p,\mr{ad}}_{\etaB}    &=&  -1.1,
  \\[1mm]
 a^T_{\kth}  &=&   8.4, & \qquad
  a^{p,\mr{ad}}_{\kth}  &=&   2.3.
 \end{array}
 \eeq
We assume $\etaB$ to be dependent on temperature and not pressure. These temperature dependencies imply that thermal gradients may induce gradients in the listed parameters, including the density and the compressibility. This leads to the appearance of the inhomogeneous acoustic body force $\fffac$ previously studied for both solute and thermal induced gradients \cite{Karlsen2016, Joergensen2021, Qiu2021}.

\subsection{Separation of length and time scales}\seclab{model_separation}
Acoustofluidic devices are typically driven at a frequency $f$ in the range from 1 to 50~MHz. The corresponding fast acoustic time scale $t$ is
 \beq{fast_time_scale}
 t = \frac{1}{\omega} = \frac{1}{2\pi f} = 3 \text{ - } 160~\SIns.
 \eeq
The time scale $\tau$ associated with the hydrodynamic and thermal flow is slower. Following the  analysis by Karlsen and Bruus \cite{Karlsen2016}, we estimate for a typical aqueous suspension in a channel of height $H = 0.5$~mm with density $\rho$, relative density difference $\hat{\rho}=0.1$ induced by gradients in either concentration or temperature, and kinematic viscosity $\nu = \eta/\rho$, the following characteristic time scales: thermal relaxation $t_\mr{therm}=H^2/\Dth$, viscous relaxation $t_\mr{visc}=H^2/\nu_0$, inertial motion $t_\mr{inert}\approx \sqrt{H/(g \hat{\rho}}))$, and steady shear motion $t_\mr{shear} \approx \nu_0/(H g \hat{\rho})$ are all in the order of $10~\SIms$. So for the slow hydrodynamic time scale $\tau$, we have
 \beq{slow_time_scale}
 \tau \approx t_\mr{therm} \approx t_\mr{visc} \approx t_\mr{inert} \approx t_\mr{shear}\approx 10~\SIms.
 \eeq
The slow thermo-hydrodynamic and the fast acoustic time scales are thus separated by 4 to 5 orders of magnitude, and we therefore solve the fast and slow dynamics separately as in Ref.~\cite{Karlsen2016}. In this work, we only study the steady limit of the slow time scale and describe any given physical field $Q_\mr{phys}$ as a sum of a steady field $Q_0$ and a time-varying acoustic field $Q_1 \ee^{-\ii \omega t}$ with a steady complex-valued amplitude $Q_1$,
\beq{Qphys}
Q_\mr{phys}(t)=Q_0+\re\big[Q_1 \ee^{-\ii \omega t}\big].
\eeq
The steady (or slowly varying) fields sets the density and compressibility, which governs the acoustic fields. Conversely, the fast acoustic time scale creates an oscillation-time-averaged acoustic body force $\fffac$ and acoustic heating-power density $\Pac$ that enter the equations of motion for the steady fields. A time-average of a product of two acoustic terms is also a steady term, as expressed by the well-known relation $\avr{\re\big[A_1 \ee^{-\ii \omega t}\big],\re\big[B_1 \ee^{-\ii \omega t}\big]}=\frac12 \re \big[ A_1 B_1^*\big]$, where the asterisk denote complex conjugation. In contrast to perturbation theory \cite{Joergensen2021}, we do not require that $Q_1$ is much smaller than $Q_0$, but we do neglect higher harmonic terms with time-dependence $e^{\pm \ii n\omega t}$, $n = 2, 3, \ldots$.

Acoustofluidic systems exhibit dynamics on two length scales, one set by the wavelength of the acoustic fields, and one set by the viscous and thermal boundary layers. The boundary conditions on the velocity field, stress, heat flux and thermal field at the fluid-solid interface results in the appearance of thermal boundary layer of width $\delt$ in both the fluid and the solid, and in a viscous boundary layer of width $\dels$ in the fluid. These boundary layers are localized near fluid-solid interfaces, and their dynamically-defined widths, jointly referred as $\delta$, are small compared to a typical device size or wavelength $d$ \cite{Karlsen2015},
 \beq{boundary_thickness}
 \dels = \sqrt{\frac{2 \nu_0}{\omega}}, \qquad
 \delt = \sqrt{\frac{2 \DthO}{(1-X)\omega}} \approx \sqrt{\frac{2 \DthO}{\omega}},
 \eeq
where $X = 0$ for fluids and $X = (\gamma-1)\frac{4\cTsqr}{3\cLsqr}  \lesssim 0.01$ for solids. Typically, $\delt \lesssim \dels \lesssim 500~\SInm$, which is more than two orders of magnitude smaller than $d\sim100~\SIum$. We introduce the usual complex wave numbers $k_s$ and $k_t$ associated with the boundary-layer widths $\dels$ and $\delt$, respectively,
 \beq{ks_kt_def}
 k_s = \frac{1+\ii }{\dels}, \qquad k_t = \frac{1+\ii }{\delt}.
 \eeq
In the following analysis, the fast acoustics fields are separated into a bulk field (superscript $d$) and boundary layer field (superscript $\delta$) that are connected by the boundary conditions.

\subsection{Boundary conditions}
In the usual Lagrangian picture \cite{Joergensen2021}, an element in an elastic solid with  equilibrium position $\sss_0$ has at time $t$ the position $\sss(\sss_0,t) = \sss_0 + \sss_1(\sss_0)\ee^{-\iot}$ and velocity  $\VVV^0 = -\pp_t\sss = \VVV _1^0(\sss_0)\: \ee^{-\ii \omega t}$ with $\VVV _1^0(\sss_0) = -\ii\omega\sss_1(\sss_0)$. On the solid-fluid interface, the no-slip and continuous stress conditions apply as in Ref.~\cite{Joergensen2021} Eqs.~(10) and~(11). The velocity of the solid wall at a given time and position must equal the Eulerian-picture fluid velocity $\vvv^\mr{fl}$,
 \beq{noslip_boundary_condition}
 \vvv^\mr{fl}(\sss_0 +\sss_1 \ee^{-\ii \omega t},t) = \VVV _1^0(\sss_0)\:\ee^{-\iot}
 = -\ii \omega \uuu_1^0(\sss_0) \:\ee^{-\iot}.
 \eeq
This boundary condition must be obeyed separately for the steady and acoustic fields (subscript 0 and 1, respectively), so a Taylor expansion yields
 \bsubal{noslip_condition}
 \eqlab{no_slip_steady}
 \vvv_0(\sss_0) &= -\avr{(\sss_1\cdot \nablabf) \vvv_1}\big|_{\sss_0},
 \\
 \eqlab{no_slip_acoust}
 \vvv_1(\sss_0) &= \VVV_1^0(\sss_0).
 \esubal
Similarly, at a given position on the fluid-solid interface with surface normal $\nnn$, the stress $\sigmabf$ must be continuous,
 \bsubal{stress_condition}
 \eqlab{stress_steady}
 \sigmabf^\mr{sl}_0(\sss_0)\cdot \nnn  & =
 \sigmabf^\mr{fl}_0(\sss_0)\cdot \nnn
 + \avr{(\sss_1\cdot\nablabf)\sigmabf^\mr{fl}_1(\sss_0)\cdot \nnn}\big|_{\sss_0},
 \\
 \eqlab{stress_acoust}
 \sigmabf^\mr{sl}_1(\sss_0) \cdot \nnn
 &=
 \sigmabf^\mr{fl}_1(\sss_0)\cdot \nnn.
 \esubal
Since the viscosity parameters $\eta$ and $\etaB$ depend on the temperature, the explicit expressions of the two stress boundary conditions contains several terms.

Following Ref.~\cite{Joergensen2021}, two sets of thermal boundary conditions must be imposed. Similar to the velocity, the temperature must be continuous across the solid-fluid interface. This condition must be obeyed separately in the steady and acoustic fields,
 \bsubal{temp_condition}
 \eqlab{temp_steady}
 \TOsl(\sss_0) &=
 \TOfl(\sss_0) + \avr{\sss_1\cdot \nablabf \TIfl}\big|_{\sss_0},
 \\
 \eqlab{temp_acoust}
 \TIsl(\sss_0) &= \TIfl(\sss_0).
 \esubal
Similar to the stress, the normal component $-\kth \nnn\cdot\nablabf T$ of the heat flux  must be continuous across the interface,
 \beq{heat_flux_boundary_condition}
 -\kthsl \nnn \cdot \nablabf T^\mr{sl}(\sss_0,t) =
 -\kthfl \nnn \cdot \nablabf T^\mr{fl}(\sss_0 +\sss_1 \ee^{-\ii \omega t},t).
 \eeq
Here and in the following, we neglect the tiny gradients in $\nnn$ and $\sss_1$. The steady and acoustic boundary conditions on the heat flux become,
 \bsubal{heatflux_condition}
 \eqlab{heatflux_zeroth}
 &-\kthOsl \nnn \cdot \nablabf \TOsl(\sss_0)
 -\avr{\kthIsl \nnn \cdot \nablabf \TIsl(\sss_0)}
 \\
 \nn
  &\qquad = -\kthOfl \nnn \cdot \nablabf \TOfl(\sss_0)
  - \avr{\kthI \nnn \cdot \nablabf \TIfl(\sss_0)}
 \\[2mm]
 \nn
 &\qquad \hspace*{5.7em} -\avr{ \sss_1 \cdot \nablabf
 \big[ \kthO \nablabf  \TIfl(\sss_0) \big]\cdot \nnn },
 \\
 \eqlab{heatflux_first}
 &-\kthOsl \nnn \cdot \nablabf \TIsl(\sss_0)
 = -\kthOfl \nnn \cdot \nablabf \TIfl(\sss_0),
 \esubal

\subsection{Range of validity of the model}
\seclab{validity_range}

We briefly discuss the range of validity imposed by the main assumptions. Firstly, in this analysis we study steady fields and acoustic fields with the actuation frequency $\omega$. So our model is only valid when these fields are much larger than the higher harmonic fields at frequencies 2$\omega$, $3\omega, \ldots$. The magnitudes $v_{0a}$, $v_{1a}$, and $v_{2a}$ of the steady $\vvvO$, the acoustic $\vvvI$, and the $2\omega$-harmonic $\vvv_2$ velocities are given in Muller and Bruus \cite{Muller2015} as,
 \beq{2ndharmonic}
 v_{0a} = \frac{Q^2 \vbc^2}{\cs},\quad  v_{1a} = Q \vbc,
 \quad  v_{2a} = \frac{Q^3 \vbc^2}{\cs},
 \eeq
where the physical velocity field corresponding to $\vvv_2$ is given as $\vvv_2^\mr{phys}=\re\big[\vvv_2  e^{-\ii 2 \omega t}\big]$. Our model is valid if $v_{1a}^2 \gg v_{2a}^2$, and this implies a limit on the acoustic energy density $\Eac \approx \frac14\rho_0 v_{1a}^2$,
\beq{Eac_limit}
\Eac \ll \frac{\rho_0 \cs^2}{4 Q^2}\approx 10^3-10^5~\SIJ\:\SIm^{-3},
\eeq
where $Q=100-1000$ is typical for acoustofluidic devices. So in systems with high $Q$ factors the higher order harmonics will be important at a lower $\Eac$.

Secondly, due to low oscillatory advection, we assume  $\nablabf \cdot (q_0 \vvv_1) \approx q_0 \nablabf \cdot \vvv_1$, where $q_0$ is a parameter of the fluid. This requires $\vert q_0 \nablabf \cdot \vvv_1 \vert \gg \vert\nablabf q_0 \cdot \vvv_1\vert$. At room temperature the validity of our theory is therefore limited by the most temperature sensitive parameter, the viscosity $\etaO$,
 \beq{T0_limit}
 \abs{\nablabf \TO } \ll \bigg|\frac{\eta_0 \kc}{(\pp^{{}}_T \eta)_{\TO}}\bigg| \approx 5000~\frac{\SIK}{\SImm}.
 \eeq
Conventional acoustofluidic systems are well within this limit $\abs{\nablabf \TO } \lesssim 50~\SIK/\SImm \ll 5000~\SIK/\SImm$.

Thirdly, the effective boundary-layer theory requires the boundary-layer width to be much smaller than the bulk wavelength, $k_0\delta \ll 1$, see \secref{model_separation}, which is true for MHz acoustics in water.

\section{Fast-time-scale acoustic fields}
\seclab{AcousticFields}
The acoustic or fast-time-scale part of thermoviscous acoustofluidics is thoroughly studied in Ref.~\cite{Joergensen2021} as the first order fields in the perturbative model. The governing equations of these fields are the same for the perturbative and the iterative model, and  therefore the theory from Ref.~\cite{Joergensen2021} can be directly applied. This is an effective theory, in which the thermal and viscous boundary layers are given analytically and incorporated in effective boundary conditions on the pressure $p_1$ and displacement field $\uuu_1$. The governing equations for the bulk fields and the effective boundary conditions on the solid-fluid boundary are given in Ref.~\cite{Joergensen2021}, Eqs.~(19), (20), (23), (24), and (34), and they are briefly summarized below.

\subsection{Governing equations in the bulk}
\seclab{acoustic_bulk}

In the bulk of the fluid, as shown in Ref.~\cite{Joergensen2021} Sec. IV A, the acoustic pressure field $p_1$ and the associated bulk velocity $\vvv_1^d$ and adiabatic temperature $\TId$ are governed by the Helmholtz equation, derived from \eqsref{govEQ_fl}{stress_fl},
\bsubal{fast_fluid}
 \eqlab{p1}
 \nabla^2 p_1 &= -\kcsqr p_1,
 \quad \kc = \frac{\omega}{c}(1+\ii\GamOcfl),
 \\
 \eqlab{v1dp}
 \vvv_1^{d,p} &= -\ii\:\frac{1-\ii\GamOcfl}{\omega\rhoO}\:\nablabf p_1,
 \\
 \eqlab{T1d}
 \TId &= (\gamma-1)\frac{\kappa_{s0}}{\alpha_{p0}}\:p_1.
 \esubal
Further, as shown in Ref.~\cite{Joergensen2021} Sec. IV B, the displacement $\uuu_1$ in the solid is governed by the temperature-dependent Cauchy equation, derived from \eqsref{govEQ_sl}{stress_sl},
 \bsubal{fast_solid}
  -\rhoO  \omega^2 \uuu_1^d &= \nablabf \cdot \sigmabf_1^{\mr{sl},d},
 \\
\sigmabf_1^{d,\mr{sl}} &= -\frac{\alfP}{\kapT} \TI \III + \taubf_1,
 \\
 \taubf_1^{d,\mr{sl}}   &= \rhoO\cTsqr \Big[\nablabf\uuu_1 + (\nablabf\uuu_1)^T\Big]\nn\\
             &\quad
             + \rhoO\big(\cLsqr-2\cTsqr\big)(\nablabf \cdot \uuu_1) \III.
 \esubal

The boundary layers at the fluid-solid interface are incorporated analytically through two effective boundary conditions. Firstly, see Ref.~\cite{Joergensen2021} Eq.~(34a), the velocity must be continuous across the interface, here imposed on $\pp_z p_1$ in the fluid,
 \bsub
 \eqlab{BCeff}
 \bal
 \pp_z p_1
 &= \ii\frac{\omega \rho_0}{1-\ii \Gams} \big(V_{1z}^0- \frac{\ii}{\ks} \divop \VVV_1^0 \big)
 -\frac{\ii}{\ks} \big(\kc^2 +\pp_z^2\big) p_1
 \nn
 \\
 \eqlab{BCeff_velocity}
 & \quad + \frac{\ii}{\kt} \frac{\alfP}{\kapT} \kOsqr \TIfldelO,\; \text{ for } z = 0,
 \eal
where $\TIfldelO$ is the boundary-layer temperature field given in the following subsection. Secondly, see Ref.~\cite{Joergensen2021} Eq.~(34b), the stress must be continuous across the interface, here imposed on $\sigmabf_1^{d,\mr{sl}}$ in the solid,
 \beq{BCeff_stress}
 \sigmabf_1^{d,\mr{sl}}\cdot \eee_z = -p_1 \eee_z +
 \ii\ks\eta_0 \Big[\vvv_{1}^{d0,\mr{sl}}  +\frac{\ii}{\omega \rho_0}\nablabf p_1 \Big].
 \eeq
 \esub
The effective boundary conditions~\eqnoref{BCeff} enable 3D simulations with a coarse mesh, because the boundary layer does not need to be resolved numerically.

\subsection{Analytical form of the boundary layers}\seclab{acoustic_bl}
The analytical solution for the boundary layers was in Ref.~\cite{Joergensen2021} used to set effective boundary conditions on the acoustic fields and the steady streaming field. Here, we also need them to derive the effective boundary conditions for the steady temperature field. The analytical solution of the temperature boundary layer in the fluid $\TIfldel$ and solid $\TIsldel$ is given  in Ref.~\cite{Joergensen2021} Eq.~(29) as,
 \bsubal{T1_delta}
 \eqlab{T1_delta_fl}
\TIfldel
 &= -\frac{\tilde{Z}}{1+ \tilde{Z}}
 \big[\TIsldO(x,y) - \TIfldO(x,y)\big] \ee^{\ii\kt^\mr{fl} z},
 \\
 \eqlab{T1_delta_sl}
 \TIsldel
 &= +\frac{1}{1+ \tilde{Z}}
 \big[\TIsldO(x,y) - \TIfldO(x,y)\big]  \ee^{-\ii\kt^\mr{sl} z},
 \\
 \eqlab{Z_def}
  \tilde{Z}  &= \frac{k^{\mr{th,sl}}_0 \kt^\sl}{k^{\mr{th,fl}}_0 \kt^\sl} =
  \sqrt{\frac{k^{\mr{th,sl}}_0 \cpO^\sl\: \rhoOsl}{k^{\mr{th,fl}}_0 \cpO^\fl\: \rhoOfl}}.
\esubal

The acoustic velocity $\vvvI$ is split into three fields, see Ref.~\cite{Joergensen2021} Eqs.~(20a): the bulk velocity $\vvv_1^{d,p}$ and the thermal boundary-layer velocity $\vvv_1^{d,T}$, both compressible gradient fields in the Helmholtz decomposition (superscript "$d$"), and the viscous boundary-layer velocity $\vvv_1^\delta$,
\beq{vvv1_decom}
\vvv_1 = \vvv_1^{d,p} + \vvv_1^{d,T}+\vvv_1^\delta
\eeq
As derived analytically in Ref.~\cite{Joergensen2021} Eqs.~(30) and (33b), $\vvvIdel$ and  $\vvv_1^{d,T}$ are given by,
 \bsubal{v1BL}
 \eqlab{v1delta}
 \vvvIdel &= \vvvIdelO(x,y)\: \ee^{\ii k_s z},
 \\
 \eqlab{v1dT}
 \vvv_1^{d,T} & =\alphapO \DthO \nablabf \TIdel = \frac{\alphapO\kthO}{\rhoO\cpO} \nablabf \TIdel.
 \esubal
These analytical expressions are used in \secref{SteadyTemp_bdr} to derive the contribution from the acoustic fields to the boundary condition of the steady thermal field.

In terms of bulk and boundary fields combined with \eqsref{eqofstate}{water_property}, the first-order density $\rhoI$, viscosity $\etaI$, and thermal conductivity $\kthI$ are written as
 \bsubalat{qIBulkBoundary}{2}
 \eqlab{rhoIBulkBoundary}
 \rhoI &= \rho_1^d + \rho_1^\delta, & 
 \rho_1^\delta &= - \rhoO\alfP \TIdel,
 \\
 \nn &&
 \rho_1^d &= \rhoO\kaps p_1,
 \\
 \eqlab{etaIBulkBoundary}
 \etaI &= \etaId + \etaIdelta, & 
 \etaIdelta &= \etaO a_\eta^{T}\alfP \TIdel,
 \\
 \nn &&
 \etaId &= \etaO a_\eta^{p,\mr{ad}}\kaps p_1,
 \\
 \eqlab{kthIBulkBoundary}
 \kthI &= \kthId + \kthIdelta\!,\;\;\;  & 
 \kthIdelta & = \kthO a_{\kth}^T \alfP \TIdel,
 \\
 \nn &&
 \kthId &= \kthO a_{\kth}^{p,\mr{ad}} \kaps p_1.
 \esubalat

\section{Slow-time-scale steady fields}
\seclab{SteadyFields}
The steady or slow-time-scale part of thermoviscous acoustofluidics contains mechanical and temperature fields. The mechanical fields are studied in Ref.~\cite{Joergensen2021}, so the equations for the displacement $\uuu_0$ in the solids and pressure $p_0$ and velocity $\vvvO$ in the fluids, can be carried over unchanged, and we just summarize the main results below. However, we need to develop the theory for the temperature field $\TO$, both its bulk part $\TOd$ and its boundary-layer part $\TOdel$, as it is not treated in Ref.~\cite{Joergensen2021}.

\subsection{Mechanical bulk and boundary-layer fields}
\seclab{SteadyMech}
As $\abs{\uuu_0}\ll d$, the steady displacement field $\uuu_0$ is to a good approximation decoupled from both the steady thermal field and the acoustic fields, and consequently
 \beq{uOequ}
 \uuu_0 = \zerovec.
 \eeq
The steady pressure $p_0$ and streaming $\vvvO$ are governed by the the steady part of \eqsref{govRho_fl}{govV_fl},
 \bsubal{v0d_gov}
 0&=\nablabf \cdot  \vvv_0^d ,\\
 0&=- \nablabf\big[p_0^d -\avr{\Lac^d} \big] - \nablabf\cdot\big[\rhoO\vvvO\vvvO] + \eta_0 \nabla^2 \vvv_0^d +\fffac^d.
\esubal
Here, the Lagrangian density $\avr{\Lac^d}$ and the acoustic body force $\fffac^d$ contain time averages $\frac12[A_1B_1^*]$ of pairs of acoustic fields $A_1$ and $B_1$, given by Ref.~\cite{Joergensen2021} Eq.~(52c),
 \bsubal{Lfacd}
 \eqlab{Lacd}
 \avr{\Lac^d} &= \frac{1}{4}\kaps\abs{p_1}^2  - \frac{1}{4}\rhoO\abs{\vvv_1^{d,p}}^2,
 \\
 \eqlab{facd}
 \fffac^d &= -\frac{1}{4}\abs{\vvv_1^{d,p}}^2\nablabf \rho_0 -\frac{1}{4}\abs{p_1}^2\nablabf\kaps
 \\ \nn
 &\quad +\bigg[1-\frac{2 a_\eta (\gamma-1)}{\beta+1} \bigg]\frac{\Gamma \omega}{c^2} \avr{\vvv_1^{d,p} p_1}
 \\ \nn
 &\quad +a_\eta \eta_0 (\gamma -1) \kcsqr \avr{\sss_1^d \cdot \nablabf \vvv_1^{d,p}}.
 \esubal
The acoustic boundary layers, are taken into account analytically, and they only appear implicitly and impose a slip velocity on the bulk streaming field given by Ref.~\cite{Joergensen2021} Eq.~(54) as,
 \bsubal{v0d_AB}
 \vvv_0^{d0} &=  (\AAA \cdot \eee_x )\eee_x + (\AAA \cdot \eee_y )\eee_y+ (\BBB \cdot \eee_z)\eee_z,
 \\
 \AAA &= - \frac{1}{2 \omega} \mr{Re}\Bigg{\{} \vvv_1^{\delta 0 *} \cdot \nablabf \bigg(\frac{1}{2} \vvv_1^{\delta 0} -\ii \VVV_1^0 \bigg) -\ii \VVV_1^{0*} \cdot \nablabf \vvv_1^{d,p} \nn
 \\
 &\quad+\bigg[ \frac{2-\ii }{2} \nablabf \cdot \vvv_1^{\delta 0 *} +\ii  \Big( \nablabf \cdot \VVV_1^{0 *} - \pp_z v_{1z}^{d *}\Big)\bigg]\vvv_1^{\delta 0 }\Bigg{\}}
 \nn\\
 &\quad + \frac{1}{2 \eta_0} \mr{Re}\bigg\{\eta_1^{d0} \vvv_1^{\delta 0*} +\frac{\delt}{\delt-\ii \dels} \eta_1^{\delta 0}  \vvv_1^{\delta 0*}  \bigg\},
 \\
 \BBB &= \frac{1}{2 \omega}  \mr{Re}\Big\{ \ii \vvv_1^{d 0 *} \cdot \nablabf \vvv_1^{d,p} \Big\},
 \\
 \vvvIdelO & = -\ii\omega \uuu_1^0 - \vvvIdO,
 \\
 \eta_1^{\delta 0} &= -\frac{\tilde{Z}}{1+ \tilde{Z}} \etaO a_\eta^T \alfP
 \big[\TIsldO - \TIfldO\big],
 \\
 \eta_1^{d 0}&= \etaO a_{\eta}^{p,\mr{ad}}\kaps p_1,
 \esubal
where the expressions  for $\vvvIdelO$, $ \eta_1^{\delta 0}$, and $\eta_1^{d 0}$ in terms of bulk fields are obtained from Eqs.~\eqnoref{noslip_boundary_condition}, \eqnoref{T1_delta_fl}, and \eqnoref{etaIBulkBoundary}.

\subsection{Steady temperature fields}
\seclab{SteadyTemp}
The steady temperature field $\TO$ is given as the time averaged terms of \eqsref{govT_sl}{govT_fl} in the solid and fluid, respectively. The time averaged terms either consist of steady fields $a_0$ or terms with time-averaged products $\avr{a_1 b_1}$ of two time-varying fields $a_1$ and $b_1$. All terms of the latter type are collected as an acoustic power $\Pac$. In the fluid, neglecting small terms by using $\vvv_0 \cdot \sigmabf_{11} \ll \avr{\vvv_1 \cdot \sigmabf_1}$, $\rho_1 \vvv_0 \ll \rho_0 \vvv_1$, $\nablabf\cdot\avr{\rho_0 \vvv_0+\rho_1 \vvv_1} = 0$, $\epsilon_{11} + \frac{1}{2}\abs{\vvv_0}^2 +\frac{1}{2}\abs{\vvv_1}^2 \ll \epsilon_0$,   $\epsilon_1^d =\cpO \TId -\frac{\alphapO \TO}{\rhoO}p_1^d=0$, $\epsilon_1^\delta =\cpO \TIdel$, and $\abs{\vvv_0\cdot\sigmabf_0}\ll\abs{ \kthO \nablabf T_0}$, the steady part of \eqref{govT_fl} becomes
 \bsubal{EC0}
 0 &= \nablabf \cdot\Big[\kthO \nablabf \TOfl\Big]
 -\cp \rho_0\vvv_0\cdot \nablabf \TOfl
 +\Pac^\fl  +P,\\
 \Pac^\fl &= \nablabf\cdot \Big[\avr{\kthI \nablabf \TIfl}
 -\avr{p_1 \vvv_1}+ \avr{\vvv_1\cdot \taubf_1}
 \nn\\
 &\qquad \qquad
 \eqlab{Pacfl}
 -\rho_0 \cpO \avr{\TIfl\vvv_1} \Big]
 - \cp \avr{\rho_1\vvv_1}\cdot \nablabf \TOfl,
 \esubal

In the solid there is no advection, and the $\TO$ part of \eqref{govT_sl} is controlled by thermal diffusion alone,
\bsubal{energy_solid}
 0 &= \nablabf\cdot(\kthO \nablabf \TOsl)+\Pac^\sl +P,
 \\
 \eqlab{Pacsl}
 \Pac^\sl &= \nablabf\cdot\avr{\kthI \nablabf \TIsl}.
\esubal
In both the solid (sl) and the fluid (fl), the temperature field and the acoustic power are separated into a boundary term ($\delta$) and a bulk term ($d$),
\beq{T_bulk_boundary}
\TO^\xl =\TO^{\xl,d}+\TO^{\xl,\delta}, \qquad \Pac = \Pac^d + \Pac^\delta.
\eeq
The boundary-layer temperature fields $\TOfldel$ and $\TOsldel$ are defined as the response to $\Pac^\delta$, and all three fields are required to go to zero far away from the boundary.

The two bulk and two boundary-layer fields are linked by the boundary conditions \eqsnoref{temp_steady}{heatflux_zeroth} at the fluid-solid interface, which impose continuity of the temperature and of the heat flux density. The first is
 \bsub
 \eqlab{ThermalBC0}
 \beq{T_bc}
 \TOfldO + \TOfldelO + \avr{ \sss_1 \cdot \nablabf \TIfl}
 =  \TOsldO + \TOsldelO,
 \eeq
and the second is
 \bal\eqlab{heat_bc}
 &\kthO \nnn \cdot \nablabf (\TOfld + \TOfldel)
   + \avr{\kthI \nnn \cdot \nablabf \TIfl }
 \nn\\
 &\hspace*{9.5em}
 + \avr{ \sss_1 \cdot \nablabf \big( \kthO \nablabf \TIfl\big) \cdot \nnn}
 \nn\\
 &= -\kthOsl \nnn \cdot \nablabf \big(\TOsld + \TOsldel\big)
 -\avr{\kthIsl \nnn \cdot \nablabf \TIsl}.
 \eal
 \esub
Thus, the steady bulk solid and fluid fields $\TOsld$ and $\TOfld$ can be matched at the interface by using the analytical form of the boundary-layer fields $\TOsldel$ and $\TOfldel$.

\subsection{Steady boundary-layer temperature fields}
\seclab{SteadyTemp_bdr}
In the fluid, the boundary-layer $\TIfldel$  is driven by $\Pac^\delta$ of \eqsref{Pacfl}{T_bulk_boundary}. We neglect the convection term  $\cpO \left(\rho_0 \vvv_0 +\avr{\rho_1 \vvv_1}\right)  \nablabf T$ in the boundary layer, because it contains only one gradient $\propto \delta^{-1}$, and thus is a factor $k\delta$ smaller than the viscous term $\nablabf\cdot\avr{\vvvI\cdot\taubfI}$ containing two gradients $\propto \delta^{-2}$. Moreover, in the boundary layer $\nablabf\cdot\big[\kth_0 \nablabf \TOfldel \big]  \approx \kth_0 \pp_z^2 \TOdel$, so the governing equation for the steady boundary-layer temperature field $\TOfldel$ therefore reduces to,
 \bsubal{T0deltaGov}
 \eqlab{EC_2ndorder_delta2}
 \kthO \pp_z^2 \TOfldel &= -\Pac^\delta
 \\
 \eqlab{Pac_delta}
 \Pac^\delta&= \nablabf\cdot \Big[ \avr{\kthI \nablabf \TI}^{\fl,\delta}
 + \avr{ \vvv_1 \cdot \taubf_1 }^{\fl,\delta}
 \\ \nn
 &
 \hspace*{3.8em} -\avr{p_1 \vvv_1}^{\fl,\delta}
 - \rho_0 \cpO \avr{\TI\vvv_1}^{\fl,\delta}\Big].
 \esubal
\bsub
The first-order boundary-layer fields are known analytically, see \secref{acoustic_bl}, so we can now analyze the four terms in $\Pac^\delta$ one by one and integrate \eqref{EC_2ndorder_delta2} once from $z=\infty$ to $z=0$ to find the normal derivative $\pp_z \TOfldel$ and twice to find the value $\TOfldel$. We describe each field at $z=0$ as a surface field (with superscript \qmarks{0}), which depends only on $x$ and $y$, multiplied by the exponential $z$ dependence given analytically in \secref{acoustic_bl}. The reduction and integration of the four terms is straightforward but tedious as shown in \appref{appA}. The normal gradient $\pp_z \TOfldel$ at the fluid-solid interface becomes \eqref{ppz_T0delta_final},
 \bal
 & \pp_z \TOfldelO =  \mr{Re}\bigg[\frac{1\!+\!\ii}{4\DthO}
 \bigg\{\frac{1\!-\!\ii}{2}\frac{\dels\omega}{\cp}\vvv_1^{\delta 0} \scap \vvv_1^{\delta 0*}
 -\frac{\delt\omega\alphapO}{\cp\rhoO} p_1^0 \TIdelOS
 \nn\\
 &\;
 -\frac{\dels}{\dels +\ii\delt}
 \Big[\delt\nablabf_\parallel \TIdelO
 \cdot\vvv_{1,\parallel}^{\delta 0*} -(1-\ii) \TIdelO v_{1,z}^{\delta 0*}\Big]
 \nn \\
 & \; - \ii \TIdelO \Big[v_{1,z}^{d,T0 *}
    + (1+\ii)  v_{1,z}^{d,p0 *} - \delt \omega\kaps p_1^{0*}\Big]
 \nn\\
 \eqlab{ppzTOdelOfl}
 &\;
 - \delt \nablabf_\parallel \TIdelO\scap
 \vvv_{1,\parallel}^{d,p0 *}
 +\delt\omega
 \frac{\kthIdeltaO \!+\! \kthIdO}{\kthO} \TIdelOS
 \bigg\}\bigg],
\eal
where all quantities are evaluated in the fluid. The boundary-layer heat flux $\kthOfl \pp_z\TOfldelO$ is dominated by the first term $\vvv_1^{\delta 0}\scap \vvv_1^{\delta 0*}$ which is a factor of $\alphapO \TO\simeq 10$ larger than the terms including the boundary layer temperature field $\TIdelO$. The two last terms including ${\kthI}$ are smaller by a factor of $\gamma-1$ and a therefore only important for gases and not liquids. The result \eqref{T0delta_final} for the boundary-layer interface temperature $\TOfldelO$ is,
 \bal
 & \TOfldelO =  \frac{\delt}{4\DthO} \mr{Re}\Big\{
 -\frac{\dels^2\omega}{2\delt\cp} \vvv_1^{\delta 0} \cdot \vvv_1^{\delta 0*}
 + \frac{\delt\omega\alphapO}{\cp\rhoO}  p_1^0 \TIdelOS
 \nn\\
 &\;
 +\ii\:\frac{\dels^2}{(\dels +\ii\delt)^2}
 \Big[\delt\nablabf_\parallel \TIdelO
 \cdot\vvv_{1,\parallel}^{\delta 0*} -(1-\ii) \TIdelO
 v_{1,z}^{\delta 0*}\Big]
 \nn
 \\ &\; - \TIdelO\Big[\frac{1-\ii}{2}  v_{1,z}^{d,T0 *}
  + (1+\ii) v_{1,z}^{d,p0 *} - \delt\omega\kaps p_1^{0*}\Big]
 \nn\\
 \eqlab{TOdelOfl}
 &\;
 + \ii\delt\nablabf_\parallel \TIdelO\cdot
 \vvv_{1,\parallel}^{d,p0 *}
 -\delt\omega\frac{(1+\ii )\kthIdeltaO + 2\kthIdO}{2\kthO} \TIdelOS
 \Big\}.
 \eal

\esub
Again, the first term $\vvv_1^{\delta 0}\scap \vvv_1^{\delta 0*}$ originating from the viscous boundary layer is the leading term.

In the solid, the boundary layer field $\TOsldel$ is governed by $P^\delta_\mr{ac}$  of \eqsref{Pacsl}{T_bulk_boundary} as,
 \beq{T0solid_delta}
 0 = -\nablabf \cdot \Big[\kthOsl \nablabf \TOsldel+\avr{\kthIsl \nablabf \TIsl}^\delta \Big],
\eeq
which, when using $\nablabf^2 \TOsldel\simeq\pp_z^2 \TOsldel$, gives the following differential equation for the boundary-layer field $\TOsldel$,
\beq{EC_2ndorder_delta_solid}
\kthO \pp_z^2 \TOsldel= -\nablabf\cdot   \avr{ \kthI \nablabf \TIsl}^\delta.
\eeq
The right-hand side is similar to $\nablabf \cdot \avr{\kthI \nablabf \TIfl}^\delta$ in the fluid boundary layer, which contributes with terms of the type $\kthI \TIdelS$ in \eqsref{ppzTOdelOfl}{TOdelOfl}. These terms in the fluid domain can be directly transferred to the solid domain, which results in the following normal heat flux and temperature in the boundary layer on  the solid side of the fluid-solid interface,
 \bsubal{solid_bl_bc}
 \eqlab{ppzT0delOsl}
 \pp_z \TOsldel &=
 \frac{1}{2\delt}\mr{Re}\bigg[(1+\ii)
 \frac{\kthIdeltaO + \kthIdO}{\kthO} \TIdelOS\bigg],
 \\
 \eqlab{T0delOsl}
 \TOsldel &= -\frac14 \mr{Re}\bigg[
\frac{(1+\ii )\kthIdeltaO + 2\kthIdO}{\kthO} \TIdelOS
 \bigg],
 \esubal
where all quantities are evaluated in the solid. For a fluid-solid interface these terms are negligible compared to the leading term in the fluid boundary layer. They can be important for certain gas-solid interfaces.

\subsection{Steady bulk temperature fields}
The steady bulk temperature field $\TOfld$ in the fluid ís governed by the long-range bulk terms of \eqref{EC0},
 \bsubal{slow_T_fluid}
 0&= -\nablabf\cdot \Big[\kthO \nablabf \TOfld \Big]-\cp\rho_0\vvv_0 \cdot \nablabf \TOfld\nn\\
 &\quad
 +\Pac^d + P,
 \\
 \Pac^d &= -\nablabf\cdot \Big[\avr{\kthId \nablabf \TId} -\avr{p_1 \vvv_1^{d,p}}
 + \avr{\vvv_1^{d,p}\cdot \taubf_1^d}\Big]
 \nn\\
 &\quad
 - \cp \avr{\rho_1^d\vvv_1^{d,p}}\cdot \nablabf \TOd .
 \esubal
Similarly, $\TOsld$  in the solid is governed by the long-range bulk terms of \eqref{energy_solid},
 \bsubal{slow_T_solid}
 0 &= - \nablabf \cdot \Big[ \kthO \nablabf \TOsld \Big] +\Pac^d+ P,
 \\
 \Pac^d &= -\nablabf\cdot\avr{\kthIsl \nablabf \TIsld}.
 \esubal
Here, $P$ is an external heat power source from fields not included in the model, such as heat generated by light absorption or by Joule heating from electric currents. The bulk fields $\TOsld$ and $\TOfld$ are connected at the fluid-solid interface by the Dirichlet and Neumann boundary conditions~\eqsnoref{T_bc}{heat_bc}.

We choose to apply the Dirichlet condition~\eqnoref{T_bc} on $\TOfld$ in the fluid and therefore write
 \bal
 \eqlab{bc_T0}
 \TOfld & = \TOsld - \TOfldelO
 \nn\\
 &\quad
 - \frac12\re\Big[ \sss_1 \cdot \nablabf \TIfldS
 - \kt^\fl (\sss_1\cdot \nnn) \TIfldelOS\Big].
 \eal
All fields on the right-hand side can be expressed in terms of bulk fields: the acoustic boundary-layer fields $\TIfldel$ and $\TIsldel$ through $\TIfld$ and $\TIsld$ by \eqref{T1_delta}, the steady boundary-layer fields $\TOfldel$ and $\TOsldel$ through $\TIfldel$, $\TIsldel$, $\TIfld$, $\TIsld$, and $p_1$ by Eqs.~\eqnoref{qIBulkBoundary}, \eqnoref{TOdelOfl}, and \eqnoref{T0delOsl}, and finally $\sss_1$ through $\uuu_1$ by the simple identification $\sss_1 = \uuu_1(\sss_0)$. Consequently, $\TOfld$ is given solely by steady and acoustic bulk fields, a crucial point in the implementation of a numerical simulation involving only bulk fields, which avoids the numerically demanding resolution of the narrow boundary layers. Note how the boundary-layer fields results in a discontinuity in the bulk temperature field, when crossing the fluid-solid interface.

Conversely, the Neumann  boundary condition~\eqnoref{heat_bc} is enforced on the temperature field $\TOsld$ in the solid. Together with the evaluation of the steady boundary layer terms in \eqsref{ppzTOdelOfl}{solid_bl_bc}, it becomes,
 \bal
 \eqlab{bc_ppzT0}
 &\kthOsl \nnn \cdot \nablabf \TOsld
 = \kthOfl \nnn \cdot \nablabf \TOfld + \kthOfl \pp_z \TOfldel
 \nn\\
 &\quad
 + \frac{1}{2}\re\Big[\kt^\fl \kthIfl \TIfldelS  - \frac{2\ii}{\delt^2}
 \kthOfl (\sss_1 \cdot \nnn)   \TIfldelS \Big].
 \eal
Similar to \eqref{bc_T0}, all fields on the right-hand side of \eqref{bc_ppzT0} can be expressed in terms of steady and acoustic bulk fields through $\TIfldel$, $\TIsldel$, $\TIfld$, $\TIsld$, and $p_1$ by Eqs.~\eqnoref{qIBulkBoundary}, \eqnoref{ppzTOdelOfl}, and \eqnoref{ppzT0delOsl}, and by using $\sss_1 = \uuu_1(\sss_0)$. Consequently, $\pp_z \TOfld$ is given solely by steady and acoustic bulk fields.

\begin{figure*}[t!]
 \centering
 \includegraphics[width=\textwidth]{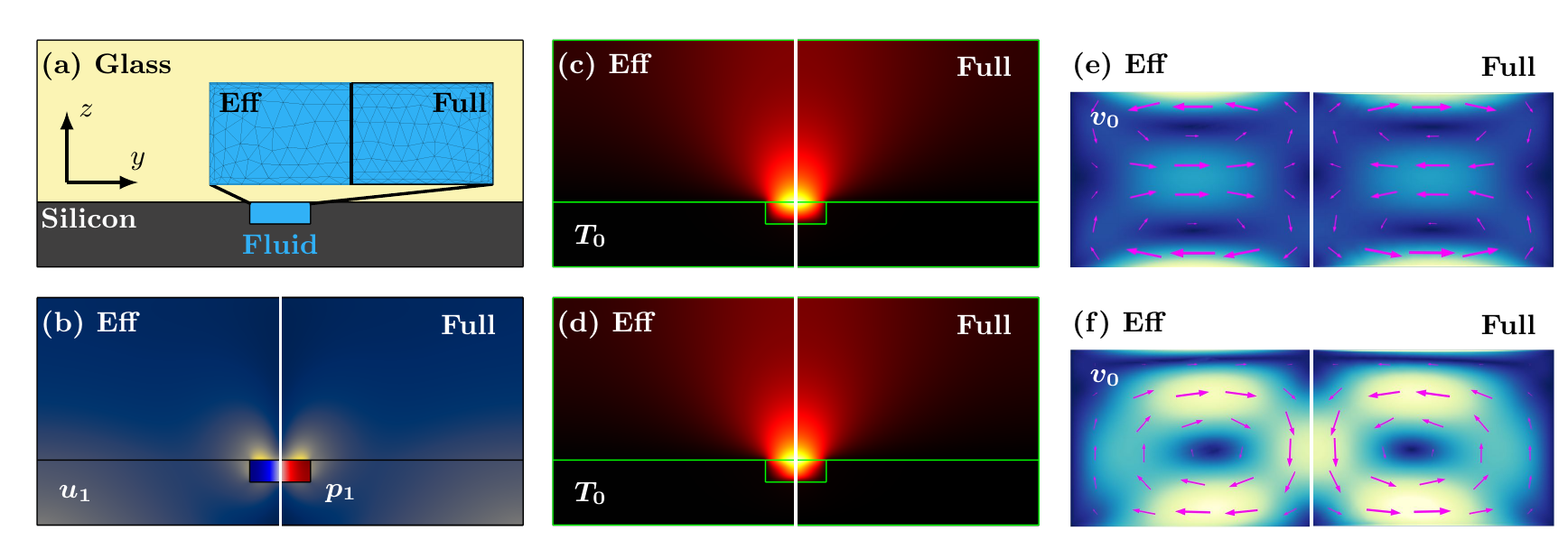}
 \caption[]{\figlab{2Dsystem}
 Comparison between simulation results of the effective model (left, Eff) and the full model (right, Full). (a) Sketch of the 2D model of the chip with silicon, fluid and glass.  (b) Color plot at the energy density $\Eac=28~\SIJ/\SIm^3$ of the displacement $\abs{\uuu_1}$ and pressure $p_1$ in the fluid. (c) Color plot at $\Eac=28~\SIJ/\SIm^3$ of the steady temperature field $\Delta T_0 = \TO-\TO^\mr{bot}$ from black (0) to yellow ($8.7~\SImK$). (d) Color plot at $\Eac=2680~\SIJ/\SIm^3$ of $\Delta\TO$ from black (0) to yellow ($230~\SImK$). (e) Vector plot at $\Eac=28~\SIJ/\SIm^3$ of the streaming $\vvv_0$ and color plot of its magnitude $v_0$ from blue (0) to yellow  ($34~\SImum/\SIs$). (f) Same as (e) but at $\Eac=2680~\SIJ/\SIm^3$ and with the color scale of $v_0$ from blue (0) to yellow ($4.0~\SImm/\SIs$).
}
 \end{figure*}

In summary, the bulk temperature fields are governed by equation \eqsref{slow_T_fluid}{slow_T_solid} together with the effective boundary conditions  \eqsnoref{bc_T0}{bc_ppzT0}, in which the boundary-layer fields are taken into account analytically and expressed in terms of bulk fields. The boundary conditions on the outer surfaces could either be a Dirichlet boundary condition, such as Peltier elements or heat sinks, a no-flux boundary condition as for an air interface, or a combination such as air cooling and solids made of glass or polymer with a thermal diffusivity similar to water.

\section{An iterative procedure to account for nonlinear effects}
\seclab{iterative}
The separation of time scales leaves us with one set of equations presented in \secref{AcousticFields} for the acoustic fields $p_1$ and $\uuu_1$, and another set presented in \secref{SteadyFields} for the steady fields $\vvv_0$, $p_0$, $\TOsl$ and $\TOfl$. These steady and acoustic fields impact each other through the temperature-dependent material parameters, the acoustic body force $\fffac^d$, the acoustic power $\Pac$, and the effective boundary conditions, in which the boundary-layer fields are taken into account analytically but appear only implicitly through expressions involving only bulk fields. As described in the following, the combined set of equations can be solved by a self-consistent iterative procedure, in which the coupled acoustic and steady fields are solved in an iterative sequence until convergence is obtained.

The steady fields $\vvv_0$, $p_0$, $\TOsl$ and $\TOfl$ are computed from the governing equations in the bulk, \eqnoref{v0d_gov} \eqnoref{slow_T_fluid} and \eqnoref{slow_T_solid} with the effective boundary conditions~\eqnoref{v0d_AB}, \eqnoref{bc_T0} and \eqnoref{bc_ppzT0}. The acoustic fields $p_1$ and $\uuu_1$ computed from the governing \eqnoref{fast_fluid} and \eqnoref{fast_solid} with the effective boundary conditions~\eqsnoref{BCeff_velocity}{BCeff_stress}.

The equations are implemented in COMSOL Multiphysics \cite{Comsol56} using the \qmarkstt{Weak Form PDE Module}, and the effective boundary conditions are set using the \qmarkstt{Dirichlet boundary condition Module} and the \qmarkstt{Weak Contribution Module}. The iterative solver is implemented using the \qmarkstt{Segregated Solver} with two steps:  \qmarkstt{Step 1} computes the steady fields $\vvv_0$, $p_0$, $\TOsl$ and $\TOfl$ based on the current value of the acoustic fields, and \qmarkstt{Step 2} computes the acoustic fields $\uuu_1$ and $p_1$ based on the current value of the steady fields. The segregated solver then runs until convergence is obtained.

The benefit of the iterative setup compared to the traditional perturbation setup \cite{Muller2012, Bach2018, Skov2019, Joergensen2021} is that nonlinear effects are included. In the steady fields there are two dominating nonlinear effects in typical microscale acoustofluidic devices: (1) Thermal convection proportional to $\vvv_0\cdot\nablabf T_0$, which dominates over thermal diffusion proportional to $\nabla^2 T_0$, when $ \abs{\vvv_0} \gtrsim \DthO/d \approx 0.3$ - $1.5~\SImm/\SIs$ for $d = 100$ - $500~\SImum$. Note that for larger systems convection becomes important at lower velocities, and in recent experimental studies this limit has been reached \cite{Qiu2021, Michel2021}. (2) Acoustic heating, which is due to the viscous dissipation $\Pac^\delta$ in the viscous boundary layer, and which may lead to temperature gradients in the bulk large enough to result in a significant acoustic body force proportional to $|p_1 |^2\nablabf \TO$ through the temperature-dependent compressibility and density, see \eqref{facd}, that drive an acoustic streaming, which at sufficiently high acoustic energy densities dominates over the usual boundary-driven Rayleigh streaming.

\section{Model validation and examples}
\seclab{ModelValidation}
In this section, we implement and validate our self-consistent iterative procedure in Comsol. We also study two specific examples of the above-mentioned nonlinear effects, which our model is able to predict.

\subsection{Example in 2D: Change of the acoustic streaming due to internal acoustic heating}
\seclab{example2D}

Our basic perturbative thermoviscous acoustofluidics model has been validated both numerically \cite{Joergensen2021} and experimentally \cite{Qiu2021}. Therefore, we here choose our first example to be a system, where we can validate numerically our model with the effective boundary conditions \eqsnoref{bc_T0}{bc_ppzT0} with a full model, where the boundary layers are fully resolved. The chosen model system, is a long straight microchannel with a rectangular cross-section, embedded in a silicon base and capped with a glass lid. In the literature, this system running with a horizontal acoustic half-wave resonance has been widely used to separate particles in a flow through device and used in various studies both experimentally \cite{Barnkob2010, Augustsson2011, Barnkob2012a, Muller2013} and numerically \cite{Muller2012, Muller2014}. Moreover, in the Letter that we published simultaneously with this work \cite{Joergensen2022}, we have provided experimental validation of the numerical model being presented below. The example aims to demonstrate three important points: (1) Validation of the effective model, (2) modeling the internal acoustic heating in an acoustofluidic chip, and (3) demonstrating nonlinear effects at high acoustic energies, effects that are further investigated by modeling and experiments in  Ref.~\cite{Joergensen2022}.

The model is a long straight silicon chip of width $W_\mr{Si}=3~\SImm$ and height $H_\mr{Si}=0.4~\SImm$, in the top of which is placed a fluid channel of width $W = 375~\SImum$ and height $H=135~\SImum$ and a capping Pyrex glass lid of height $H_\mr{Py}=1~\SImm$, see \figref{2Dsystem}(a). On the bottom edge, following Ref.~\cite{Joergensen2021}, the actuation  is set to be $\uuu_1^\mr{bot}(y) = \frac{2 d_0}{W} y\, \eee_z$, and the temperature is $T_0^\mr{bot} = 25~\SICel$.

In \figref{2Dsystem}(a) is shown the coarse mesh of the effective model (422 elements) and the fine mesh (8362 elements) of the boundary-layer-resolving full model that are needed to fulfil a mesh-convergence criterium of an $L_2$-norm \cite{Muller2012} for the steaming velocity $\vvvO$ below 1\% for $\Eac=28~\SIJ/\SIm^3$. The good agreement between the two models is shown in \figref{2Dsystem}(b)-(f) by the color plots of the resulting steady and acoustic fields computed from the effective (left side) and full (right side) model at both a low and a high acoustic energy density of $\Eac=28~\SIJ/\SIm^3$ and $2680~\SIJ/\SIm^3$. Both models show how the well-known four-roll Rayleigh streaming pattern at the low $\Eac$ change into a two-roll pattern at high $\Eac$, a clear display of the nonlinear effect arising from the boundary layers, but nevertheless included in the effective model of the bulk fields. The relative deviation between the two models in the computed values of the resonance frequency and Q-factor of the 2-MHz half-wave resonance mode is less than 0.1\%.

\begin{figure}[t!]
 \centering
 \includegraphics[width=\columnwidth]{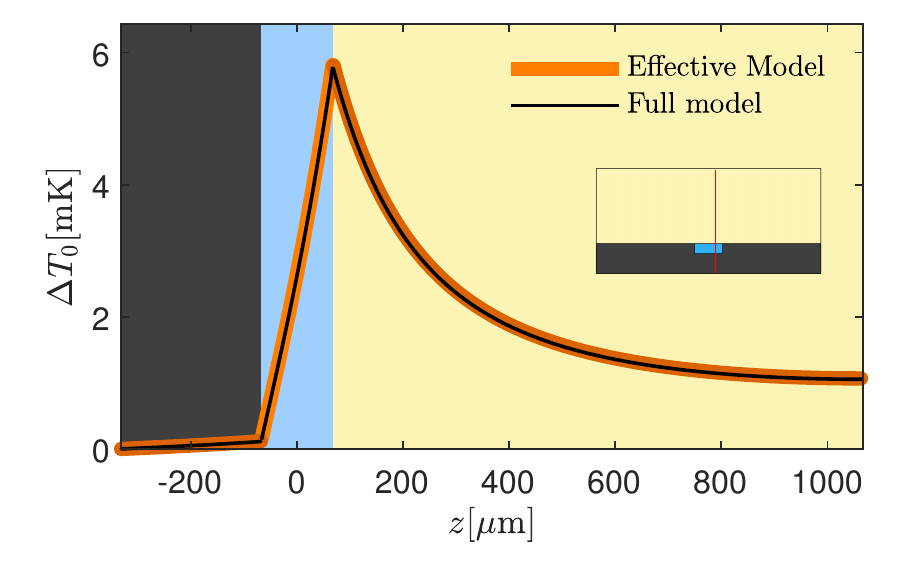}
 \caption[]{\figlab{valid_T0}
 Line plot at $\Eac=28~\SIJ/\SIm^3$ of the simulated steady temperature $\Delta\TO$
 in the effective and full model along the vertical red line $y = \frac14 W_\mr{fl}$ shown in the inset. The corresponding color plot of $\Delta\TO$ is shown in \figref{2Dsystem}(c).
 }
 \end{figure}

The effective boundary condition for the streaming velocity $\vvvO$ was already validated in Ref.~\cite{Joergensen2021}, so here we thus just need  to validate  the effective boundary conditions~\eqsnoref{slow_T_fluid}{slow_T_solid} on the steady temperature field $\TO$. This in done in \figref{valid_T0}, showing excellent quantitative agreement between line plots of $\TO$ for the full and the effective model.

\begin{figure}[t!]
 \centering
 \includegraphics[width=\columnwidth]{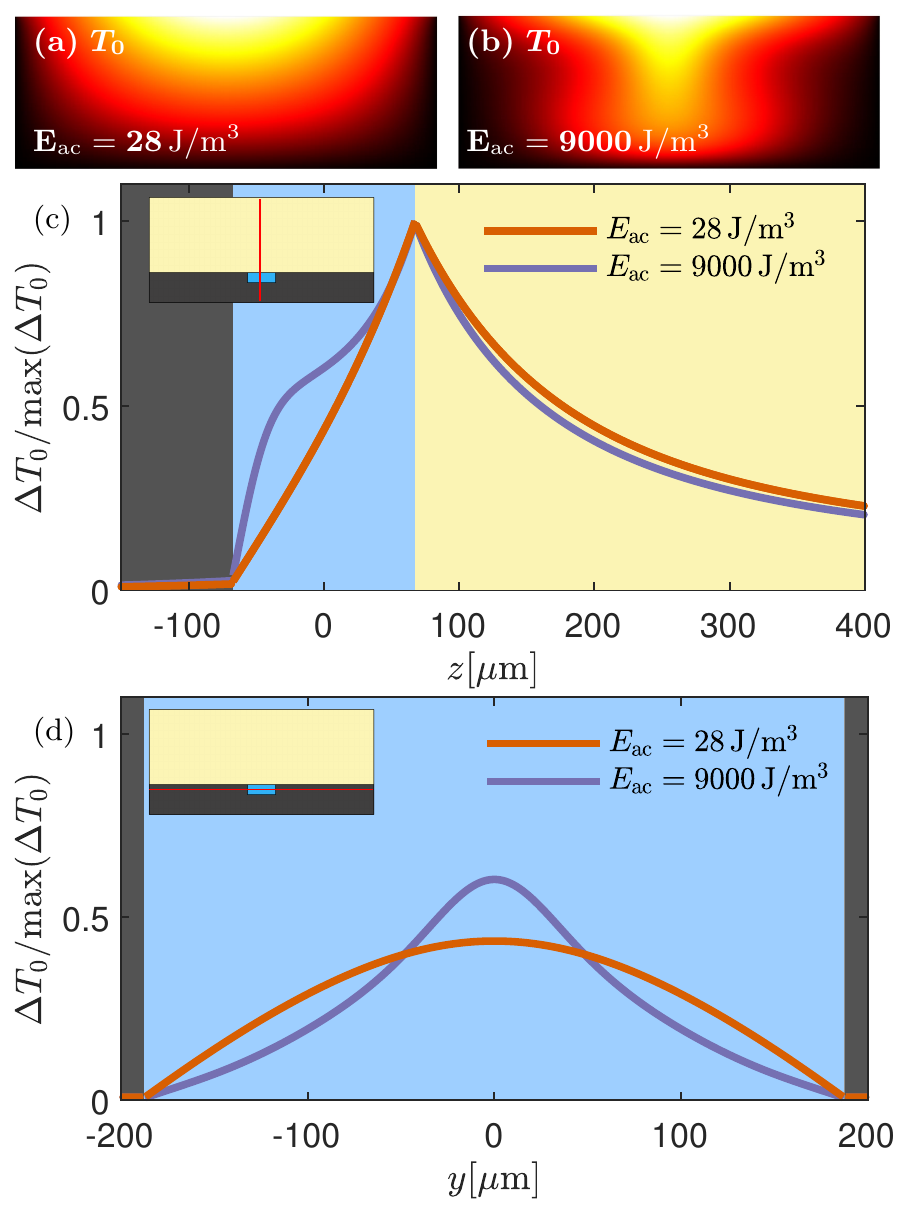}
 \caption[]{\figlab{convection}
 (a) Color plot from 0 (black) to $8.7~\SImK$ (yellow) at $\Eac=28~\SIJ/\SIm^3$ of $\Delta \TO$ from \figref{2Dsystem}(c) zoomed in on the fluid domain. (b) Same as (a) but for  $\Eac=9000~\SIJ/\SIm^3$ and a color scale from $0$ (black) to $2375~\SImK$ (yellow). (c) Line plot of the normalized temperature rise $\Delta\TO/\mr{max}(\Delta\TO)$ along the vertical line at $y=0$ shown in the inset. (d) Same as (c) but along the horizontal line through the center of the microchannel shown in the inset.}
 \end{figure}

We end the example by discussing the physics of the transition from the linear case with four flow rolls to the nonlinear case with two flow rolls. At the low acoustic energy $\Eac=28~\SIJ/\SIm^3$, the acoustic pressure $p_1$ and displacement $\uuu_1$ field, as well as the steady streaming field $\vvv_0$ are shown in \figref{2Dsystem}(b). The source of the spatial inhomogeneities in the steady temperature field $\TO$ in the fluid is the heat generation due to friction in the viscous boundary layer in the fluid at the top and bottom of the channel, and the different heat fluxes resulting from the relatively small values of the heat conductivity of water and glass compared to the large one of silicon. The latter ensures efficient transport of heat away from the bottom edge of the channel. Consequently, heating only occurs at the top of the channel near the glass lid. In \figref{convection}(a)-(b), the resulting temperature fields are shown for a low  $\Eac=28~\SIJ/\SIm^3$ and high $\Eac=9000~\SIJ/\SIm^3$ energy density, respectively. In both cases, the temperature is clearly larger at the center of the top edge of the channel. However, for the high-$\Eac$ case, the increased acoustic streaming is distorting the temperature field, as it induces a downward heat convection, which stretches the high-temperature region along a larger portion of the vertical center axis.
The temperature boundary condition~\eqnoref{bc_T0} results in nearly equal bulk and boundary layer temperature fields at the fluid-solid interface, $\TOfldO \approx \TOfldelO$, so the gradients in the temperature field are governed by the effective boundary condition on the heat flux~\eqnoref{bc_ppzT0}.

The streaming fields for $\Eac = 28~\SIJ/\SIm^3$ and $\Eac = 2680~\SIJ/\SIm^3$ are shown in \figref{2Dsystem}(e)-(f). First we can see that the full and effective model results in the same streaming field. Secondly, it is clear that at low acoustic energies the streaming is dominated by boundary driven streaming and at high acoustic energy it is dominated by the bulk-driven streaming induced by the acoustic-body-force \eqref{facd}. When the gradients in density and compressibility is created due to temperature gradients, and neglecting Eckart streaming, the acoustic body force is given as
 \beq{fac_T}
 \fffac^d =
 -\frac{1}{4}\Big(\abs{p_1}^2 \pp^{{}}_T \kappa_{s,0} +\abs{\vvv_1}^2 \pp^{{}}_T \rho_0 \Big) \nablabf \TO.
 \eeq
The temperature gradient $\nablabf \TO$ and $\vert p_1^2\vert$ both scale with the acoustic energy density $\Eac$, so the acoustic body force $\fffac^d$, and thus the streaming, scales with  $\Eac^2$. In comparison, the boundary-driven Rayleigh streaming scales with $\Eac$. Consequently, the bulk-driven streaming driven by $\fffac^d$ will become dominant at sufficiently high $\Eac$. We study further the nonlinear behavior and transition both numerically and experimentally in the Letter~\cite{Joergensen2022} published simultaneously with this work.

Finally, we note that it is important that the device consists of a silicon base with a glass lid and not a pure glass chip, because the asymmetry of the thermal field due to the widely different thermal conductivities in the two materials results in a skew-angled body force which enables a strong thermoacoustic streaming. The modeling of the transition into the nonlinear regime has not been captured by the previous perturbation models in the literature \cite{Muller2014, Bach2018, Joergensen2021}, because it requires a nonlinear model such as the one presented here.

\subsection{Example in 3D: Nonlinear thermoacoustic streaming driven by absorption of light}
\seclab{example3D}
The effective boundary conditions enables 3D simulations that combined with the iterative solver makes it possible to investigate highly nonlinear effects in a 3D system. As an example, we choose the system, in which we previously studied both experimentally and numerically the thermoacoustic streaming induced by the temperature gradient \cite{Qiu2021}. In that study, the applied perturbative model was at its limit of validity because of the high streaming velocity. Therefore, {we here use the iterative model to examine the nonlinear effects in this system, specifically the impact of advection in the system at high streaming velocities. The example serves to demonstrate the ability to make 3D models with an effective iterative model, and to study the nonlinear thermoviscous effects due to thermal convection in 3D.

\begin{figure}[t!]
 \centering
 \includegraphics[width=\columnwidth]{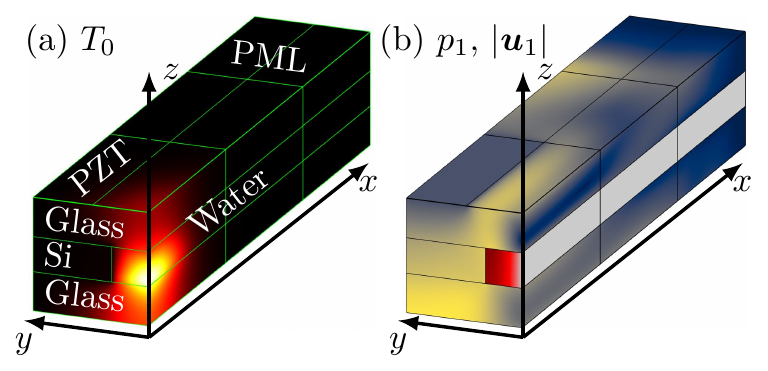}
 \caption[]{\figlab{3D_T0_p1}
Simulation of $\TO$, $\pI$, and $|\uuu_1|$ in a quarter of the glass-Si-glass system. (a) Color plot of $\TO$ from 20 (black) to $20.8~\SICel$ (yellow) due to the absorption of light from an LED with $P = 5$~mW. (b) Color plot of the corresponding acoustic displacement $\abs{\uuu_1}$ from $0$ (blue) to $18~\SInm$ (yellow) in the solid, and the acoustic pressure $p_1$ in the water-filled $0.76\times0.36~\SImm^2$ microchannel from $0$ (gray) to $1.2~\SIMPa$ (red).
}
 \end{figure}

The system is a glass-silicon-glass chip with a long rectangular water-filled channel of width $W_\mr{fl}=760~\SImum$ and height $H_\mr{fl}=360~\SImum$, such that the top and bottom of the fluidic channel is in contact with glass and the sides are in contact with the silicon wafer. The chip is actuated anti-symmetrically around the $xz$-plane and symmetrically around the $yz$-plane at a frequency $f_0 =0.96~\SIMHz$ which excites the half-wave resonance in the width of the channel. Dye has been added to the water to absorb the light from a light-emitting diode (LED). The absorbed light heats up the water and induces a temperature gradient and thus the acoustic body force $\fffac^d$ \eqref{fac_T} in the bulk. As a result, high streaming velocities and thermal convection appears. In contrast to the 2D example of \secref{example2D}, we keep the acoustic energy density $\Eac$ constant in the 3D example, and only vary the power of the LED. The acoustic body force $\fffac^d$, and thus the streaming velocity $\vvvO$, depends linearly on $\nablabf T_0$, and therefore depends linearly on the LED power as long as thermal convection is negligible.

For the numerical model we are using symmetry planes and perfectly matched layers (PML) to reduce the size of the 3D model. The LED is placed in the center of the channel, so both the  $yz$-plane at $x=0$ and the $xz$-plane at $y=0$ are symmetry planes. On the $yz$-plane all fields (steady and acoustic fields) are symmetric, whereas on the $xz$-plane the steady fields are symmetric, while the acoustic pressure $p_1$ and $y$-component of the displacement field $\uuu$ are anti-symmetric, the $x$- and $z$-component of the displacement field are symmetric. The two symmetry planes allows to only solve a quarter of the system as shown in \figref{3D_T0_p1}. The PML layer is used to dampen waves traveling along the $x$-axis away from the center, and it allows us to restrict the computational domain to the region closest to the LED spot. Further details on the implementation of the PML layer and boundary conditions on the symmetry plane can be found in the Supplemental Material~\footnote{See Supplemental Material at \url{http://bruus-lab.dk/files/Joergensen_nonperturb_theory_Suppl.pdf} for details about the implementation of the symmetry planes and PML in the 3D model.}.

The actuation is implemented as a displacement on the glass-lid, which sets up an acoustic field with an energy density $\Eac=150~\SIJ/\SIm^3$ at $x=0$, and the LED is modeled to be a Gaussian beam centered at $x=y=0$ and with a width of $650~\SImum$. With a LED power of $P=5~\SImW$ the resulting steady temperature field $\TO$ and acoustic pressure $p_1$ and displacement field $\uuu_1$ are shown in \figref{3D_T0_p1}. $\TO$ is strongest at the bottom of the fluidic channel, because the light is absorbed there and the silicon wafers keeps the sides of the channel cold be transporting the heat to a heat sink.

When the LED is off, the streaming is dominated by the boundary-driven streaming, but when it is on, the streaming is dominated by the acoustic body force $\fffac^d$. The transition from boundary- to bulk-driven streaming is thoroughly studied in Ref.~\cite{Qiu2021}. The resulting streaming for three different LED powers are shown in \figref{3D_streaming}. Here, the solution in quarter channel has been mirrored in the two symmetry planes to obtain the streaming flow in the full channel. In \figref{3D_streaming}(a) is shown the classical boundary-driven Rayleigh streaming for zero LED power $P$. In this case the streaming pattern contains four characteristic 2D streaming rolls in the  $yz$-plane  with  almost no  flow in the $x$-direction. In \figref{3D_streaming}(b) is shown the streaming for

\begin{widetext}
\mbox{}\noindent
\begin{figure}[h]
\centering
\includegraphics[width=\textwidth]{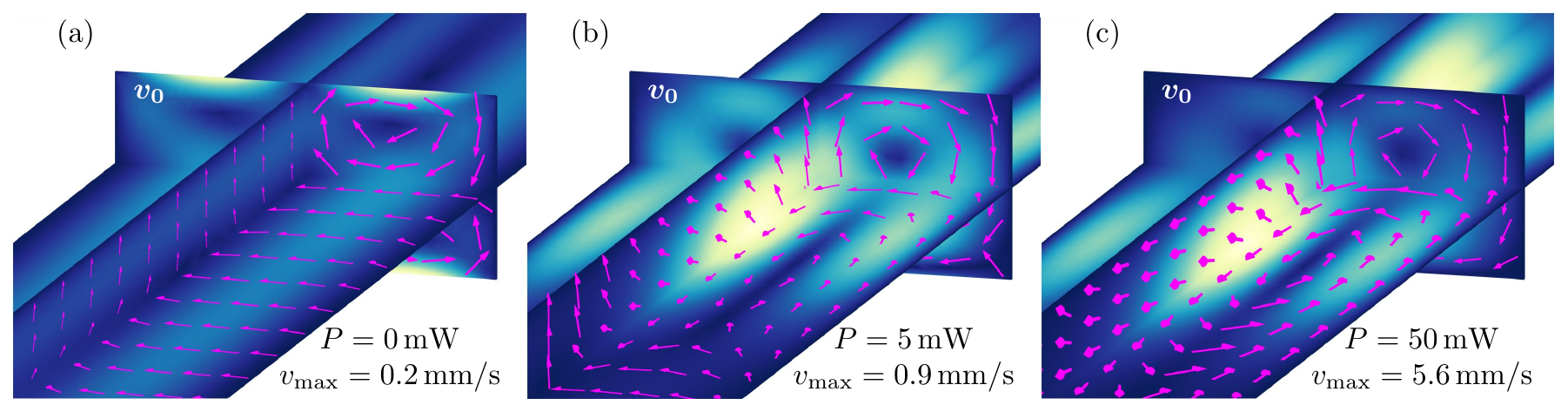}
\caption{\figlab{3D_streaming}
Simulated streaming $\vvv_0$ at acoustic energy density $\Eac=150~\SIJ/\SIm^3$ for the LED power $P = 0$, $5$, and $50$~mW, respectively. The color plots from $0~\SImm/\SIs$ (blue) to $v_\mr{max}$ (yellow) are the in-plane velocity of the respective planes, on the $yz$-plane it is $(v_{0,y}^2+v_{0,z}^2)^{1/2}$, and likewise for the $xy$- and $xz$-planes. All arrows are unit vectors showing the direction of $\vvv_0$.  (a) $\vvv_0$ for $P = 0$~mW showing the usual four boundary-driven streaming rolls with $v_\mr{max} = 0.2$~mm/s. (b) $\vvv_0$ for $P = 5$~mW showing a slightly dominant thermoacoustic streaming flow with $v_\mr{max} = 0.9$~mm/s driven by the acoustic body force $\fffac^d$ \eqnoref{fac_T} in the bulk. (c) $\vvv_0$ for $P = 50$~mW completely dominated by the fast streaming flow with $v_\mr{max}=5.6~\SImm/\SIs$ driven by the the acoustic body force $\fffac^d$ in the bulk.}
\end{figure}
\end{widetext}

\begin{figure}[t!]
\centering
\includegraphics[width=\columnwidth]{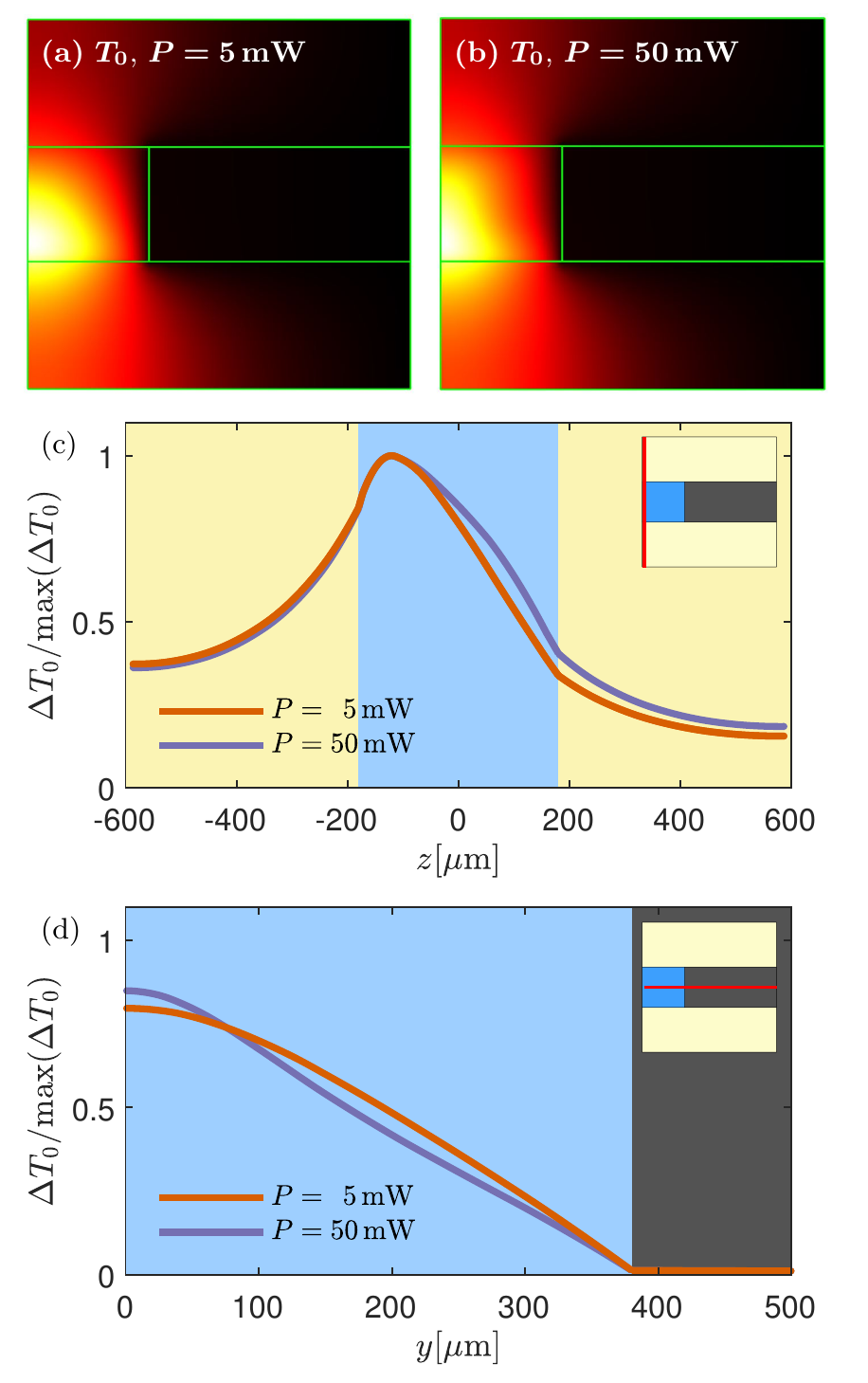}
\caption[]{\figlab{3D_diff_conv}
Convection due to high streaming velocities. (a) shows the temperature field in the $yz$-plane at $x=0$ generated by the light absorption from a LED of power $P=5~\SImW$ ranging from $\TO =20.0~\SIC$ (black) to $\TO=20.8~\SIC$ (yellow). (b) Same as (a) but for $P=50~\SImW$ and a color-range from $\TO =20.0~\SICel$ (black) to $\TO=27.3~\SICel$ (yellow). (c) Shows a line plot of the two normalized temperature fields along a line at $x=y=0$ shown in the inset. (d) Shows a line plot of the two normalized temperature fields along a line at $x=z=0$ shown in the inset. The difference in the two temperature fields are due to convection.
}
\end{figure}

\noindent
moderate LED power $P=5~\SImW$ with a maximum velocity of $0.9~\SImm/\SIs$, which recovers the 3D flow pattern driven by the bulk acoustic body-force $\fffac^d$ as observed in Ref.~\cite{Qiu2021}. In \figref{3D_streaming}(c) is shown the streaming for high LED power $P=50~\SImW$ with a maximum velocity of $5.6 ~\SImm/\SIs$. This pattern looks like the one for $P=5$~mW, but is slightly deformed due to changes in $\nablabf\TO$ and thus in $\fffac^d$ due to nonlinear thermal convection.

The temperature fields for $P=5$ and $50~\SImW$ are shown in \figref{3D_diff_conv}(a,b) in the $yz$-plane at $x = 0$. In the case of $P=50~\SImW$, the streaming-induced convection has stretched the temperature field up along the center axis and thereby altering $\TO$ and $\fffac^d$. This stretching reduces the temperature gradient and $\fffac$ along $z$, and therefore leads to the reduction of $\vvv_0$ in the vertical $yz$-plane relative to the one in the horizontal $xy$-plane seen when comparing \figref{3D_streaming}(b) to (c). The dependence of the $\Delta\TO$-profile on the LED power $P$ is quantified by the line plots of the temperature along the vertical line $y=x=0$ in \figref{3D_diff_conv}(c) and the horizontal line $x=z=0$ in \figref{3D_diff_conv}(d). These line plots show the same tendency as was observed in the 2D-example in \figref{convection}.

When simulating convection-diffusion processes, the numerical mesh needs to satisfy the stability condition $h_\mr{mesh}< 2 \DthO/v_0$ on the numerical P\'eclet number, which restricts the size $h_\mr{mesh}$ of the mesh elements. In this system, with $\DthO \approx 2\times10^{-7}~\SIm^2/\SIs$ and $v_0 = 5.6~\SImm/\SIs$, we find $h_\mr{mesh} < 40~\SImum$.
Consequently, in systems with a high streaming velocity, a fine mesh is required in the bulk, which quickly can make numerical 3D simulation computationally expensive. It is possible to mathematically stabilize the diffusion-advection equations which can enable simulations with a coarser mesh, but this we have not yet implemented in our simulation.

\section{Conclusions}
\seclab{conclusions}
We have presented an effective nonlinear model for thermoviscous acoustofluidics, which enables simulations of high acoustic energies in 3D. The model differs from previous acoustofluidic models \cite{Bach2018, Muller2014} on two main points: (1) it contains an effective boundary condition for the steady temperature field, which enables 3D simulations of acoustic heating in thermoviscous acoustofluidics, and (2) it relies on an iterative solver, which incorporate nonlinear effects, and thus allows simulations of higher acoustic energies than models based on perturbation theory.

To validate the model and to demonstrate its potential, we firstly presented a 2D example of a widely used rectangular channel was modeled in \secref{example2D}. In \figref{2Dsystem}, the effective model was validated against a full iterative model, and the internal acoustic heating due to friction was shown in \figref{2Dsystem}(b)-(c) to be of the order $\SImK$. Secondly, the capability of simulating nonlinear effects in 3D systems was demonstrated in \secref{example3D}, an example showing the importance of convective heat transport in a acoustofluidic device with externally controlled temperature gradients. We have presented experimental validation of the nonlinear model as well as further experimental and numerical studies of the transition from perturbative to nonperturbative behavior as a function of $\Eac$ around $500~\SIJ/\SIm^3$ in the Letter~\cite{Joergensen2022} published simultaneously with this work. We note that $\Eac \gtrsim 500~\SIJ/\SIm^3$ can easily be obtained in standard acoustofluidic devices, where $\Eac \approx 10 - 50~\SIJ/\SIm^3\:\big[U_\mr{pp}/(1~\SIV)\big]^2$ has been reported in the literature, $U_\mr{pp}$ being the applied voltage on the piezoelectric transducer \cite{Barnkob2010, Augustsson2011, Barnkob2012, Muller2013}.

In many applications of acoustofluidic devices, as high a throughput as possible is desired. Generally, a higher acoustic energy will allow for such a higher throughput. The presented iterative model allows simulations of higher acoustic energies and will likely contribute to an increased understanding of nonlinear effects in acoustofluidics and to an improved design capability of acoustofluidic devices with a higher throughput.

\begin{acknowledgments}
This work was supported by Independent Research Fund Denmark, Natural Sciences (Grant No.~8021-00310B).
\end{acknowledgments}

\appendix
\section{Reduction and integration}\seclab{appA}
In this appendix, we present the mathematical steps going from \eqref{T0deltaGov} to \eqsref{ppzTOdelOfl}{TOdelOfl} for the heat flux and the temperature in the fluid at the fluid-solid interface. Beginning with \eqref{T0deltaGov}, but suppressing the superscript \qmarks{fl} for simplicity, we have
 \bal\eqlab{EC_2ndorder_delta2_app}
 &\kthO \pp_z^2 \TOdel=
 \\
 \nn
 &-\nablabf\scap \Big[  \avr{\kthI \nablabf \TI}^\delta
 \!+\avr{\vvv_1 \scap \taubf_1 }^\delta
 \!-\avr{p_1 \vvv_1}^\delta \!-\rho_0 \cpO\avr{\TI\vvv_1}^\delta\Big].
 \eal
First, the four terms on the right-hand side are evaluated and reduced one by one. Then, they are integrated with respect to $z$, once to find $-\kthO \pp_z \TOdel$, and twice to find $\TOdel$, which both are needed for the boundary conditions in \eqsref{T_bc}{heat_bc}. Similarly, we repeatedly use in the following that gradient terms are dominated by $z$ derivatives of boundary-layer fields $\TIdel$, $\vvv_1^\delta$, and $\vvv_1^{d,T}$, as each such derivative results in a factor $(k_c\delta)^{-1} \gg 1$. We also note that $\avr{(ip_1)p_1} =\avr{(i\TI)\TI} = 0$, and another helpful relation is found in Ref.~\cite{Joergensen2021} Eq.~(33a),
 \beq{div_v1}
 \nablabf\cdot\vvv_1 = i(1-\ii \Gamma_s)\omega \kapsO\:p_1 -\ii \omega \alpha_{p0}\:\TIdel,
 \eeq
revealing that $\nablabf\cdot\vvv_1$ depends not only on the bulk pressure $p_1$ but also on the boundary-layer temperature field $\TIdel$. Using this insight together with the exponentially decaying boundary-layer fields from \eqref{v1BL}, we find for the pressure-generated power,
 \bal
 \nablabf &\cdot \big[\avr{p_1 \vvv_1}^\delta \big]
 = \big[\avr{\nablabf p_1 \cdot \vvv_1^\delta} + \avr{ p_1 \nablabf\cdot \vvv_1}\big]
 \nn\\
 \eqlab{PowerPressure}
 &\approx \frac{\omega}{2} \bigg[\rhoO\re\Big\{ \ii \vvv_1^{d,p} \cdot\vvv_1^{\delta *} \Big\}
 +\alphapO \re\Big\{ \ii p_1  \TIdelS \Big\}\bigg].
\eal
The first term, being of the order $\omega \rhoO v_1^2 \approx \omega \kapsO p_1^2$, turns out to be the dominant term. Likewise, for the heat-generated power, we find that
 \bal
 \nablabf & \cdot \big[\cpO \rhoO \avr{\TI\vvv_1}^\delta \big]
\approx \cp \rhoO \big[\avr{\nablabf \TIdel \cdot \vvvI}
 + \avr{\TIdel (i\omega\kapsO p_1)}\big]
 \nn\\
 \eqlab{PowerHeat}
 &\approx \frac{\cpO \rhoO}{2} \mr{Re}
 \Big\{k_t \TIdel v_{1z}^{\delta*} -\ii \omega \kaps \TIdel p_1^*\Big\}.
\eal

In the stress-generated power $\nablabf\cdot\avr{\vvv_1 \cdot \tau_1}^\delta$, we keep only terms  $\pp_z \vvv_1^\delta$, each producing a factor of $(k_c \delta)^{-1} \gg 1$. Thus, $\vvv_1 \cdot \taubf_1 \approx \etaO \big[v_{1z} \big(\pp_z \vvv_1^\delta\big) + \big(\vvv_1\cdot \pp_z \vvv_1^\delta\big) \een_z\big]$ gives,
 \bal
 \nablabf \cdot  & \avr{\vvv_1 \cdot\taubf_1}^\delta
 \approx \etaO \nablabf\cdot
 \avr{v_{1z} \big(\pp_z \vvv_1^\delta\big)} + \etaO\pp_z\avr{\vvvI\cdot\big(\pp_z \vvv_1^\delta\big)}
 \nn \\
 & \approx \eta_0\Big[\avr{\big(\pp_z v_{1z}^\delta\big)^2}
      + \avr{\big|\pp_z \vvv_1^\delta\big|^2}
      + \avr{\vvvI\cdot\ppsqr_z \vvv_1^\delta}\Big]
 \nn \\
 & = \frac{\eta_0}{2}\re\Big[|k_s|^2 v_{1z}^\delta v_{1z}^{\delta*}
      + |k_s|^2 \vvvI^\delta \cdot\vvvI^{\delta*}
      - (k_s^*)^2 \vvvI\cdot\vvv_1^{\delta*}\Big]
 \nn \\
 \eqlab{PowerStress}
 &= \frac{\rhoO \omega}{2} \re\big[
 v_{1z}^\delta v_{1z}^{\delta*}
      + \vvvI^\delta \cdot\vvvI^{\delta*}
 +\ii  \vvv_1^{d,p} \cdot\vvv_1^{\delta*}
 \big].
\eal
Here, we have used that $\re\big[i\vvv_1^\delta\cdot v_1^{\delta*}\big] = 0$, and that $k_s = (1+\ii )\delsInv$ implies the relations $(k^*_ s)^2 = -\ii 2\delssqrInv$, $|k_s|^2 = 2 \delssqrInv$, and $2\eta_0 \delssqrInv = \rhoO\omega$.

The last term is the power generated by thermal conduction, which only contains the thermal boundary layer characterized by the wave number $k_t = (1+\ii ) \deltInv$, \eqref{T1_delta},
 \bal
 \nablabf \cdot\avr{ &\kthI \nablabf \TI}^\delta
\approx \avr{(\pp_z \kthI)\:\pp_z \TIdel} + \avr{\kthI\ppsqr_z \TIdel}
 \nn\\
 & = \frac{1}{2}\re \Big[\kt \kt^* \kthIdelta \TIdelS - \big(\kthId+ \kthIdelta\big) \big(\kt^*\big)^2 \TIdelS\Big]
 \nn\\
 \eqlab{PowerThermCond}
 &= \frac{1}{\delt^2}\re \Big[(1+\ii ) \kthIdelta \TIdelS + \ii\kthId \TIdelS\Big]
\eal
For water, this term is a factor $(\gamma -1) a_k \alphapO \TO  \approx 10^{-2}$ smaller than $\omega \kapsO p_1^2$, as can be seen by using $\deltsqrInv \kthI \approx \omega \rhoO \cp \kthI/\kthO \approx \omega \rhoO \cp a_k \alphapO \TI = a_k \omega \alphapO^2 \TO p_1$ and $\TI \approx (\gamma-1) (\kapsO/\alphapO)\:p_1$. So the power generated by thermal conduction can be neglected in fluids, but it may be important for gases.
Inserting the power contributions \eqnoref{PowerPressure}-\eqnoref{PowerThermCond} into \eqref{EC_2ndorder_delta2_app}, we arrive at the expression
 \bal\eqlab{EC_2ndorder_delta_terms}
 &\kthO \pp_z^2 \TOdel =
 \nn\\
 &\quad -\frac{\omega\rhoO }{2} \mr{Re}\Big\{\vvv_1^\delta\cdot \vvv_1^{\delta*} \Big\} + \frac{\omega \alphapO}{2} \re\Big\{\ii p_1  \TIdelS \Big\}
 \nn\\
 &\quad
 +  \frac{\cp \rhoO}{2}\mr{Re} \Big\{ \nablabf \TIdel \cdot \big(\vvv_1^{\delta*}+\vvv_1^{d,T*}+\vvv_1^{d,p*}\big)
 -\ii \omega \kaps \TIdel p_1^*\Big\}
 \nn\\
 &\quad
 -\frac{\omega\cp \rhoO}{\kthO}
 \re \Big\{ \left(1+\ii \right) \kthIdelta \TIdelS + \ii\kthId  \TIdelS\Big\},
 \eal
where the first term is the leading term, which originates from the viscous boundary layer.
This expression is now integrated from $z=\infty$ to $z=0$ once to obtain the heat flux and twice to obtain the boundary-layer temperature at the interface.
The fields in the boundary layer are well approximated by surface fields that does not depend on the normal coordinate $z$ but only on the in-plane coordinates $x$ and $y$, according to the following separations,
 \bal\eqlab{taylor}
 p_1^d  &= p_1^{d0}(x,y),
 &
 \TIfldel &= \TIfldelO(x,y)  r(z),
 \nn\\
 \vvv_1^{d,p}  &=  \vvv_1^{d0,p}(x,y),
 &
  \TIsldel &= \TIsldelO(x,y)  r^*(z),
 \nn\\
 \vvv_1^\delta  &=\vvv_1^{\delta 0}(x,y) q(z),\!\!\!
 &
\vvv_1^{d,T} & = \alphapO \DthO \nablabf \big[\TIfldelO(x,y) r(z)\big],
 \nn\\
 q(z) &= e^{i \ks z},
 &
 r(z) &= e^{i \kt z}.
 \eal
Inserting this into \eqref{EC_2ndorder_delta_terms}, we obtain,
 \bal
 & \kthO \pp_z^2 \TOdel =  \frac12 \omega \rhoO \mr{Re}\Big\{-\vvv_1^{\delta 0} \cdot \vvv_1^{\delta 0*} q q^* +\ii \frac{\alphapO}{\rhoO}  p_1^0 \TIdelOS r^*
 \nn\\
 &\quad
 +\frac{\cp}{\omega} \Big[\nablabf_\parallel \TIdelO
 \cdot\vvv_{1,\parallel}^{\delta 0*}q^* +\ii\kt \TIdelO
 \big(v_{1,z}^{\delta 0*} q^* + v_{1,z}^{d,T0 *} r^*\big)\Big]r
 \nn\\
 &\quad
 +  \frac{\cp}{\omega} \Big[\nablabf_\parallel \TIdelO\cdot
 \vvv_{1,\parallel}^{d,p0 *}+ \ii\kt \TIdelO  v_{1,z}^{d,p0 *}
 -\ii \omega\kaps \TIdelO p_1^{0*}\Big]r
 \nn\\
 \eqlab{partial_II_T0delta}
 &\quad
 -\frac{\cp}{\kthO} \Big[(1+\ii )\kthIdeltaO r +\ii\kthIdO \Big] r^* \TIdelOS
 \Big\}.
\eal
When integrating \eqref{partial_II_T0delta} with respect to $z$, the $xy$-dependent surface fields (superscript 0) move outside the integral. Using the procedure of Ref.~\cite{Bach2018}, we introduce the integrals $\intn ab$ of the integrand $a(z)\,b(z)^*$, where $a(z)$ and $b(z)$ are any of the functions $1$, $q(z)$, and $r(z)$,
 \bal
 \eqlab{def_intn}
 \intn ab &=  \int^{z}\!\dm z_n \int^{z_n}\!\dm z_{n-1} \ldots \int^{z_2}\!\dm z_1\:
 a(z_1)\: b(z_1)^* \bigg|^{{}}_{z=0},
 \nn
 \\[-1mm]
 \intn ab & \propto \delta^n,
 \text{ with }\, \delta = \dels,\delt \text{ and }
 n = 1, 2, 3, \ldots.
 \eal
Integrating \eqref{partial_II_T0delta} once with respect to $z$ thus gives
 \bal
 & \pp_z \TOdel =  \frac{1}{2\DthO} \mr{Re}\Big\{
 -\frac{\omega}{\cp}\vvv_1^{\delta 0} \cdot \vvv_1^{\delta 0*} \intI qq
 +\ii \frac{\omega\alphapO}{\cp\rhoO}  p_1^0 \TIdelOS \intI 1r
 \nn\\
 &\quad
 +\Big[\nablabf_\parallel \TIdelO
 \cdot\vvv_{1,\parallel}^{\delta 0*} +\ii\kt \TIdelO
 v_{1,z}^{\delta 0*}\Big] \intI rq
 + \ii\kt \TIdelO v_{1,z}^{d,T0 *} \intI rr
 \nn\\
 &\quad
 + \Big[\nablabf_\parallel \TIdelO\cdot
 \vvv_{1,\parallel}^{d,p0 *}+ \ii\kt \TIdelO  v_{1,z}^{d,p0 *}
 -\ii \omega\kaps \TIdelO p_1^{0*}\Big] \intI r1
 \nn\\
 \eqlab{partial_I_T0delta_IabI}
 &\quad
 -\frac{\omega}{\kthO}
 \Big[(1+\ii )\kthIdeltaO \intI rr +\ii\kthIdO \intI 1r \Big] \TIdelOS
 \Big\}.
\eal
Inserting here the values of $\intI ab$ given by
 \bal\eqlab{IntValI}
 \intI r1 &= -\frac{1+\ii}{2}\delt, &
 \intI rr &= -\frac{1}{2}\delt, & I_{ba}^{(n)} = \Big[I_{ab}^{(n)}\Big]^*,
 \nn\\
 \intI rq &= -\frac{1+\ii}{2}\:\frac{\dels\delt}{\dels +\ii\delt}, &
 \intI qq &= -\frac{1}{2}\dels, &
 \eal
leads to $\pp_z \TOfldelO$ at the fluid-solid interface,
 \bal
 & \pp_z \TOfldelO =  \mr{Re}\bigg[\frac{1\!+\!\ii}{4\DthO}
 \bigg\{\frac{1\!-\!\ii}{2}\frac{\dels\omega}{\cp}\vvv_1^{\delta 0} \scap \vvv_1^{\delta 0*}
 -\frac{\delt\omega\alphapO}{\cp\rhoO} p_1^0 \TIdelOS
 \nn\\
 &\;
 -\frac{\dels}{\dels +\ii\delt}
 \Big[\delt\nablabf_\parallel \TIdelO
 \cdot\vvv_{1,\parallel}^{\delta 0*} -(1-\ii) \TIdelO v_{1,z}^{\delta 0*}\Big]
 \nn \\
 & \; - \ii \TIdelO \Big[v_{1,z}^{d,T0 *}
    + (1+\ii)  v_{1,z}^{d,p0 *} - \delt \omega\kaps p_1^{0*}\Big]
 \nn\\
 \eqlab{ppz_T0delta_final}
 &\;
 - \delt \nablabf_\parallel \TIdelO\scap
 \vvv_{1,\parallel}^{d,p0 *}
 +\delt\omega
 \frac{\kthIdeltaO \!+\! \kthIdO}{\kthO} \TIdelOS
 \bigg\}\bigg].
\eal

To obtain the temperature $\TOdelO$ at the fluid-solid interface, we integrate \eqref{partial_II_T0delta} twice with respect to $z$. This is easily done by changing $\intI ab$ to $\intII ab$ in \eqref{partial_I_T0delta_IabI} followed by insertion of the values
 \bal\eqlab{IntValII}
 \intII r1 &= \frac{\ii}{2}\delt^2, &
 \intII rr &= \frac{1}{4}\delt^2,\nn\\
 \intII rq &= \frac{\ii}{2}\:\frac{\dels^2\delt^2}{(\dels +\ii\delt)^2}, &
 \intII qq &= \frac{1}{4}\dels^2.
 \eal
The result for $\TOfldelO$ becomes
 \bal
 & \TOfldelO =  \frac{\delt}{4\DthO} \mr{Re}\Big\{
 -\frac{\dels^2\omega}{2\delt\cp} \vvv_1^{\delta 0} \cdot \vvv_1^{\delta 0*}
 + \frac{\delt\omega\alphapO}{\cp\rhoO}  p_1^0 \TIdelOS
 \nn\\
 &\;
 +\ii\:\frac{\dels^2}{(\dels +\ii\delt)^2}
 \Big[\delt\nablabf_\parallel \TIdelO
 \cdot\vvv_{1,\parallel}^{\delta 0*} -(1-\ii) \TIdelO
 v_{1,z}^{\delta 0*}\Big]
 \nn
 \\ &\; - \TIdelO\Big[\frac{1-\ii}{2}  v_{1,z}^{d,T0 *}
  + (1+\ii) v_{1,z}^{d,p0 *} - \delt\omega\kaps p_1^{0*}\Big]
 \nn\\
 \eqlab{T0delta_final}
 &\;
 + \ii\delt\nablabf_\parallel \TIdelO\cdot
 \vvv_{1,\parallel}^{d,p0 *}
 -\delt\omega\frac{(1+\ii )\kthIdeltaO + 2\kthIdO}{2\kthO} \TIdelOS
 \Big\}.
\eal

Again the first term is the leading term that originates from the viscous boundary layer.

%
%


%

\end{document}